\documentclass{aa}  

\usepackage[normalem]{ulem}
\usepackage{amsmath}
\usepackage{graphicx}
\usepackage{txfonts}
\usepackage{xcolor}
\usepackage{comment}
\usepackage{multirow}

\begin{document} 

\title{HD\,7977 and its possible influence on Solar System bodies.}

\titlerunning{HD\,7977 influence on Solar System bodies.}

\authorrunning{P. A. Dybczyński et al.}

\author{Piotr A. Dybczyński
          \inst{1}\fnmsep\thanks{\email{dybol@amu.edu.pl}}
          \and
        Małgorzata Królikowska
          \inst{2}\and
        Przemysław Bartczak  
          \inst{1}\fnmsep\inst{3}\and
        Edyta Podlewska-Gaca
          \inst{1}\and
        Krzysztof Kamiński
          \inst{1}\and
        Jakub Tokarek 
          \inst{1}\and
        Krzysztof Langner
          \inst{1}\fnmsep\inst{4}\fnmsep\inst{6}\and
        Jos de Bruijne
          \inst{5}
        }

   \institute{Astronomical Observatory Institute, Faculty of Physics, Adam Mickiewicz University, Słoneczna 36, PL60-283 Poznań, Poland
    \and
   Centrum Badań Kosmicznych Polskiej Akademii Nauk (CBK PAN),
Bartycka 18A, 00-716 Warszawa, Poland
    \and 
     Instituto Universitario de Física Aplicada a las Ciencias y las Tecnologías (IUFACyT). Universidad de Alicante, Ctra. San Vicente del Raspeig s/n. 03690 San Vicente del Raspeig (Alicante, Spain)
    \and
     Istituto di Fisica Applicata “Nello Carrara” (IFAC-CNR), Sesto Fiorentino 50019, Italy   
    \and
    European Space Agency (ESA), European Space Research and Technology Centre (ESTEC), Keplerlaan 1, 2201 AZ Noordwijk, The Netherlands
    \and
    Dipartimento di Fisica, Universita di Padova, Via Marzolo 8, 35131 Padova, Italy}
   \date{Received XXXXXX; accepted XXXXX}

  \abstract
   {In the latest {\it Gaia} third data release one can find extremely small proper motion components for the star HD\,7977. This, together with the radial velocity measurement lead to the conclusion that this star passed very close to the Sun in the recent past.}
   {Such a very close approach of a one solar mass star must have resulted in noticeable changes in the motion of all Solar System bodies, especially those on less tight orbits, namely long-period comets (LPCs) and  Transneptunian objects (TNOs). We estimate and discuss these effects.} 
   {Our current knowledge on the Solar Neighbourhood found in the latest {\it Gaia} catalogues allowed us to perform numerical integrations and prepare a list of potential stellar perturbers of LPCs. We use this list, made available in the StePPeD database. To study the past motion of LPCs under the simultaneous action of the Galactic potential and passing stars, we use precise original cometary orbits taken from the current CODE catalogue. }
   {We examine the reliability of the extremely small proper motion of HD\,7977 concluding that this star can be an unresolved binary but, according to the astrometry covering more than a century, the current {\it Gaia} data cannot be ruled out. We present the parameters of a very close passage of this star near the Sun. We also show examples of the strong influence of this passage on the past motion of some LPCs. We also discuss the possible influence of this perturber on other Solar System bodies.}
   {It is possible that 2.47 Myr ago the one solar mass star HD\,7977 passed as close as 1000 au from the Sun. Such an event constitutes a kind of dynamical horizon for all studies of the past Solar System bodies' dynamics.}

\keywords{comets:general -- stars:kinematics and dynamics  -- (Galaxy:) solar neighbourhood}
   \maketitle

\section{Introduction}\label{sec:intr}

For many years, we have searched for stars that approach the Sun and therefore can perturb orbits of long-period comets (LPCs); see, for example, \cite{dyb-kan:1999,dyb-hab3:2006}. In \cite{rita-pad-magda:2020}, we introduced a publicly available Stellar Potential Perturbers Database (StePPeD)\footnote{\url{https://pad2.astro.amu.edu.pl/StePPeD}} containing data on potential stellar perturbers of LPC motion. This database had several revisions and is currently based on \textit{ Gaia} Early Data Release 3 \citep{GaiaEDR3-summary:2021, Dyb-Ber-StePPeD:2022}. Similar results, also based on \textit{ Gaia} EDR3, have also been published by \cite{Bob-Baj:2022} and \cite{Bailer-Jones:2022}. Recently, \cite{Raymond_et_al:2023} published an analysis on how an extremely close stellar passage can change the motion of Solar System bodies.

\cite{First-stars:2020} announced for the first time that the star \object{HD 7977} could be an important perturber for some observed LPCs based on  {\it Gaia} Data Release 2 \citep{Gaia-DR2:2018}. During the StePPeD database update to version 3.2, \cite{Dyb-Ber-StePPeD:2022} found that according to {\it Gaia} Early Data Release~3  \citep{GaiaEDR3-summary:2021} this star (designated as P0230 in StePPeD) nominally passed 2.47 Myr ago as close to the Sun as 0.014 pc ($\sim$3,000\,au). This result means that HD\,7977 might  have been an important perturber of the motion of most (if not all) LPCs and should be taken into account in all long-term dynamical investigations of the past motion of the Solar System bodies. Some examples of such strong interactions have recently been presented by \cite{dyb-kroli:2022}. 

This paper is organised as follows. In the next section, we summarise our current knowledge about the star HD\,7977 on the basis of all available results from the {\it Gaia} mission. This includes astrometry, distance estimates, and radial velocity measurements. Section \ref{sec:our_data} describes all sources of data that we use in our calculations. It contains both a discussion of historical data and a presentation of our own astrometry. The influence of the HD\,7977 data uncertainties on the results presented is discussed in Section \ref{sec:HD7977_uncertainties_vs_close_pass}. We also note the existence of several other potential stellar perturbers in Section \ref{sect-other-perturbers}.   The potential effect of the passage of  star HD 7977 very close to the Sun on the motion of LPCs is described in Section \ref{sec:LPCs_past_motion}. 
Some estimates of such an influence on other Solar System bodies are presented in Section \ref{sec:other-bodies_past_motion}. Our conclusions and prospects for future studies are placed in Section \ref{sec:conclusions}. This paper is accompanied by  three Appendices, which contain a detailed inspection of various current {\it Gaia} results for HD\,7977.

\section{What do we know today about HD\,7977 from the \textit{Gaia}  mission results?}\label{sec:comment_on_gaia}

This section, together with the Appendices \ref{Appendix-astrometry}, \ref{Appendix-distances}, and \ref{Appendix-spectra}, provides a thorough inspection of the various {\it Gaia} DR3 data for HD\,7977, focussing on a quality assessment of the data and the possibility that the star is an unresolved binary. We find several independent indications that hint at the possibility that HD\,7977 is an unresolved binary, namely a poor astrometric fit (Section~\ref{subsec:astrometry}), scatter in radial velocity measurements (Subsection~\ref{subsec:radial_velocity}), and spectral peculiarities (Appendix~\ref{Appendix-spectra}). 

None of the hints in isolation is sufficiently convincing, but the fact that the three hints come from three independent data sets adds weight to the unresolved binary hypothesis. The final confirmation or rejection of this hypothesis requires the publication of {\it Gaia} DR4. We recall that the fact that the star is not listed as a nonsingle star in {\it Gaia} DR3 is in no way a proof that the object is a single star. 

The $\sim$800,000 binary systems that are published in {\it Gaia} DR3 \citep{GaiaColl_Arenou_et_al:2023} are a small subset, heavily impacted by filtering and therefore heavily biased, of all binaries that {\it Gaia} has detected and that are currently being processed for publication in {\it Gaia} DR4. 

\subsection{Astrometry}\label{subsec:astrometry}

The {\it Gaia} astrometry reported in the {\it Gaia} catalogue(s) has tiny formal uncertainties, but several caveats apply. First, it is important to realise that the various {\it Gaia} data releases are not independent: the astrometry reported in {\it Gaia} DR2 \citep{Gaia-DR2:2018} is based on observations collected during the first 22 months of the mission, whereas the astrometry reported in {\it Gaia} DR3 \citep{Gaia-DR3-release:2022} is based on observations collected during the first 34 months of the mission (of which the first 22 months obviously overlap with {\it Gaia} DR2). In general, therefore, {\it Gaia} DR2 can be considered to be superseded by {\it Gaia} DR3 since it is based not only on an extended set of observations but also on improved data reductions (for example, matching of observations to sources) and calibrations \citep{2021A&A...649A...2L}.

The astrometry reported in the main {\it Gaia} DR3 source table reflects the formal parameters and formal uncertainties resulting from fitting a single-star motion model to the individual observations. The single-star assumption is clearly strong and can lead to major astrometric model errors that are not included in the formal uncertainties reported in the catalogue if the object is an unresolved binary.

Each astrometric fit is therefore accompanied by a goodness-of-fit parameter, the renormalised unit-weight error (ruwe\footnote{ https://gea.esac.esa.int/archive/documentation/GDR2/Gaia\_archive\\
/chap\_datamodel/sec\_dm\_main\_tables/ssec\_dm\_ruwe.html}), which quantifies how well the individual observation residuals of the single-star fit conform to white noise. In other words, large values of ruwe indicate a poor astrometric fit as, for instance, caused by binarity. Although a canonical threshold for acceptable astrometry was ${\rm ruwe} < 1.4$ for {\it Gaia} DR2, the value of {\it Gaia} DR3 that denotes the limit above which sources have significant astrometric deviations inconsistent with single source astrometry is 1.25 \citep{2020MNRAS.495..321P}. HD\,7977 has ${\rm ruwe} = 2.01$, well above this threshold, suggesting that the object is an astrometric binary. 

The scrupulous discussion of the \textit{Gaia} astrometric data and their quality is placed in Appendix \ref{Appendix-astrometry}. In conclusion, formal astrometric uncertainties do not tell the full story. Most importantly, they do not include model errors, possibly up to the level of a fraction of a milliarcsecond or more, which would be present if the object were an astrometric binary, as suggested by various quality measures. Moreover, the formal uncertainties are underestimated by up to a factor of 2 or more. One possible interpretation of the proper motion evolution seen between the three {\it Gaia} data releases is that it is explained by the additional months of data that have been added in the astrometric fit of each release, together with the aforementioned model error. In other words: neither the DR1, nor the DR2, nor the DR3 proper motion of HD\,7977 are reliable and the astrometric model error is (likely) orders of magnitude larger than the formal uncertainties reported in DR3.

\subsection{Distance}\label{subsec:distance}

The {\it Gaia} DR3 parallax equals $13.212 \pm 0.032$~mas (with a formal signal-to-noise ratio of 410). As discussed before, the formal uncertainty is very likely underestimated by a factor of 2 or more, even apart from possible model errors should the source be an astrometric binary. Despite this error underestimation, the relative parallax error is smaller than 1\%. This, in turn, justifies the use of the inverse of the measured parallax as an essentially unbiased distance estimate \citep{2018A&A...616A...9L}. This yields a distance estimate of $75.690 \pm 0.185$~pc. The associated absolute magnitude in the $G$ band equals
\begin{align}
M_G &= 8.891319 - 0.005 - 5.0\cdot({\rm log}_{10}(75.690)-1.0)\nonumber\\
&= 4.49 \pm 0.01~{\rm mag}\nonumber 
\end{align}
using an extinction value of ${\rm ag\_gspphot} = 0.005$~mag (see below).

The {\it Gaia} parallaxes are known to suffer from a small systematic bias that depends on magnitude, colour, and sky position \citep{2021A&A...649A...4L}. For HD\,7977, the recommended bias correction -- to be subtracted from the measured parallax -- equals $-0.024$~mas so that the 'debiased' distance estimate becomes $75.551 \pm 0.185$~pc. The bias correction itself has an RMS scatter of $\sim$0.015~mas. Since the bias estimate, its uncertainty, and the formal parallax uncertainty are all of the same order of magnitude, it is not trivial to conclude whether applying the correction makes sense or not, and hence we do not apply it. Several other distance estimates can be found in the literature; see the Appendix \ref{Appendix-distances}.

\subsection{Radial velocity}\label{subsec:radial_velocity}

The radial velocity of HD\,7977 reported in {\it Gaia} DR3 equals $26.76 \pm 0.21$~km~s$^{-1}$. Similarly to the case of astrometry (Section~\ref{subsec:astrometry}), the formal uncertainties of the radial velocities are also not perfectly estimated. Figure~8 in \cite{2023A&A...674A...5K} suggests that the uncertainty is underestimated by a factor $\sim$1.5--1.75 (HD\,7977 has ${\rm grvs\_mag} = 8.229$~mag). Two elements impact the absolute value of the radial velocity, and hence the traceback time of the encounter: the radial-velocity zero point and the stellar gravitational redshift. Equation~5 in \cite{2023A&A...674A...5K} suggests that the zero-point offset of the {\it Gaia} DR3 radial velocities is negligible at 8$^{\rm th}$ magnitude. The stellar gravitational redshift for this source is estimated to be ${\rm gravredshift\_flame} = 0.5624 \pm 0.012$~km~s$^{-1}$ \citep[for details, see Section 3.3.1 in][]{Fouesneau_et_al:2023}. A detailed discussion on the \textit{Gaia} spectral data for HD\,7977 can be found in Appendix \ref{Appendix-spectra}.

\section{How we completed the data for our calculations}
\label{sec:our_data}

\begin{table*}
        \caption{Astrometry of HD\,7977 = {\it Gaia} DR3 510911618569239040 in the last two {\it Gaia} data releases.}
        \label{gaia_pos_plx}
\setlength{\tabcolsep}{8.0pt}{
        \begin{tabular}{c c c c c c c} 
        \hline
          & Right Ascension & RA error $\cdot$ cos Dec & Declination & Dec error & Parallax & Plx error  \\
           & [deg] & [deg] & [deg] & [deg] & [mas] & [mas] \\
        \hline  
        DR2 (ICRS, 2015.5) & 20.13165220556 & $6.36\cdot10^{-9}$ & +61.88252167035 & $7.25\cdot10^{-9}$ & 13.2030 & 0.0376 \\
        DR3 (ICRS, 2016.0) & 20.13165220666 & $5.44\cdot10^{-9}$ & +61.88252164550 & $6.56\cdot10^{-9}$ & 13.2118 & 0.0322 \\
        \hline
 
        \hline
        \end{tabular}
        }
\end{table*}

Despite the possible unrecognised multiple nature of HD\,7977, while waiting for more detailed data, we now have to base our calculations on data on the bright component that has been recorded in stellar catalogues for over a hundred years.

According to our search, the star in question was mentioned for the first time in the catalogue published by \cite{Oeltzen:1842} as a star number 1383. Later, it was included by Argelander in the Bonner Durchmusterung \citep{Argelander-BD:1903} and designated as BD+61 250. Since then, this star has been included in almost all catalogues of reference stars covering this part of the northern sky, since it is relatively bright ($\sim$9{~mag}). The position of BD+61~250 has been for example absolutely determined by Krueger during the Astronomische Gesellschaft observing campaign in 1869-1880 in Helsigfors and Gotha for the declination zone from +55 to +65 zone \citep{Krueger:1890}. In the Henry Draper Catalogue \citep{HD-cat:1918} it is named HD\,7977 and we will use this name throughout the rest of the text. More information can be found in the SIMBAD database\footnote{\url{https://simbad.cds.unistra.fr/simbad/sim-id?Ident=hd+7977}} \citep{SIMBAD-article:2000}.

To calculate the past motion of this star through the Galaxy and obtain the parameters of its close passage near the Sun we need its current position, proper motion, parallax, and radial velocity. These data allow us to calculate the six-dimensional starting point for the backward numerical integration of its motion. 

\subsection{\textit{Gaia} position and parallax}\label{sec:position}

As the source of the astrometric position and the parallax value for HD\,7977, we took the most recent data provided by the {\it Gaia} mission, namely the {\it Gaia} EDR3 catalogue \citep{GaiaEDR3-summary:2021}, which was later incorporated into the complete {\it Gaia} DR3 release \citep{Gaia-DR3-release:2022}. Both position and parallax are consistent between {\it Gaia} DR2 and DR3 releases, as shown in Table \ref{gaia_pos_plx}. Only in the case of declination do some small discrepancies exist. The situation with the proper motion of HD\,7977 is much more complicated.

\subsection{Proper motion problem}\label{sec:proper-motion}

In searching for stars closely passing near the Sun, it is obvious that we are interested mainly in stars with the smallest observed proper motion. Because of the well-known statistical property:  "large proper motion means small distance" stars with an infinitesimal proper motion were excluded from many observational projects dedicated for parallax or radial velocity measurements, as noted in \cite{dyb-kwiat:2003}. HD\,7977 is such a kind of star, with a very small proper motion and, therefore, difficult to measure.

Several historical and contemporary determinations of HD\,7977 proper motion components are collected in Table \ref{tab:proper-motion}. To our knowledge, the first reliable value of its proper motion can be found in the AGK3 catalogue \cite{AGK3-1975}. It was obtained by a comparison of the astrometric position obtained in 1928.9 for AGK2 catalogue \citep{AGK2-1951} and observations made in 1956.73. Unfortunately, the authors did not present an uncertainty of this proper motion value. As a resort, we use mean proper motion errors in AGK3 as estimated by \cite{Schwan:1985} that are relatively large. It may be unexpected that, based on the same historical position obtained in 1928.9,  the authors of the SAO catalogue \citep{SAO:1966} obtained a completely different proper motion of HD\,7977. \cite{Bucciarelli-AGK3U:1992} published an updated version of the AGK3 catalogue, again with the different proper motion of this star; see Table \ref{tab:proper-motion}.

An independent line of proper motion measurements for HD\,7977 is based on the absolute position determination carried out at the Vatican Observatory in the course of the Carte du Ciel project \citep{Vatican_zone:1914}. The star in question was recorded on two photographic plates in 1903. An attempt to determine proper motions based on these old photographic observations has been made by \cite{Urban-Vatican-Vizier:1996} who re-reduced original Vatican plates using the published plate measurements. These positions were used to produce the ACRS catalogue \citep{ACRS:1991}. Based on the first edition of the Tycho catalogue \citep{TYC-HIP-Cats:1997} and several old catalogue positions, \cite{ACT:1998} calculated new proper motions for HD\,7977 which is included in the ACT catalogue. Another group of catalogues we examined consists of PPM \citep{ppm_cat:1988} and PPMXL \citep{PPMXL:2010}. Probably the most reliable catalogue values before the {\it Gaia} era can be found in the Tycho-2 catalogue \citep{tycho2-cat:2000}. 

It was a big surprise when proper motions well below 1~mas/yr has been presented in the {\it Gaia} DR2 \citep{Gaia-DR2:2018}. In {\it Gaia} EDR3 \citep{GaiaEDR3-summary:2021} even a much smaller value for the proper motion in the Right Ascension is presented. See Section \ref{sec:comment_on_gaia} for a possible explanation and interpretation of {\it Gaia} DR2 and EDR3 proper motion inconsistency for HD\,7977.

\begin{table}
        \caption{Proper motion measurements for HD\,7977. All numbers are taken from electronic versions of the quoted catalogues included in the Vizier database \citep{vizier}. }
        \label{tab:proper-motion}
\setlength{\tabcolsep}{7.8pt}{
        \begin{tabular}{@{}llr@{$\>\pm\>$}lr@{$\>\pm\>$}l@{}}
        \hline
         & & \multicolumn{2}{c}{$\mu_{\alpha}\cdot \cos(\delta)$}         & \multicolumn{2}{c}{$\mu_{\delta}$} \\
        Year & Source \& epoch   & \multicolumn{2}{c}{[mas/yr]} & \multicolumn{2}{c}{[mas/yr]} \\
        \hline  
    1966 & SAO B1950                & $-32.5$ &     8.0                   & +18.0     &      14.0 \\
    1975 & AGK3 B1950               & +4   &   4.4\tablefootmark{a} & $-6$    &     9.3\tablefootmark{a} \\
    1988 & PPM J2000                & +2.1    &     2.8                   & $-7.0$    &       5.6  \\
    1991 & ACRS B1950               & +2.0    &     2.3\tablefootmark{b}  &  $-4.9$  &       4.9\tablefootmark{b}  \\
    1992 & AGK3U B1950              & +0.2    &     4.71                  & +1.9      &       4.71 \\
    1997 & ACT J2000                & $-4.52$ &     1.8                   & $-4.7$    &       2.28 \\
    2000 & Tycho-2 J2000            & $-3.5$  &     2.0                   & $-1.8$    &       1.9  \\
    2010 & PPMXL 1991.0             & $-3.3$  &     1.8                   & $-0.8$    &       1.8  \\
    2018 & \textit{Gaia} DR2 2015.5 & $0.559$ &     0.019                 & $0.014$   &       0.046  \\
    2022 & \textit{Gaia} DR3 2016.0 & $0.144$ &     0.011                 & $0.010$   &       0.034  \\
        \hline
        \end{tabular}
        \\
        \tablefoottext{a}{For AGK3 proper motion errors we use the mean catalogue error of the proper motion estimated by \cite{Schwan:1985}.}\\
        \tablefoottext{b}{In ACRS proper motion errors for HD\,7977 are not given so we apply mean catalogue errors estimated by \citep{Corbin:1989}.}
        }
\end{table}

\subsubsection{Our own astrometry}\label{sec:astrometry-our}

Due to the great spread of published HD\,7977's proper motion components,  caused by different combinations of positions, large possible errors of historical measurements, and different precession/nutation standards and time bases, we decided here to perform our own astrometric position estimations. For this purpose, we used the following data sources: two photographic plate measurements from the Vatican Observatory survey \citep{Vatican_zone:1914} obtained in 1903, two scans of photographic plates from the Palomar Observatory Sky Survey~I obtained in 1954 \citep{Pennington:1993}, one scan of photographic plate from the Palomar Quick~V survey obtained in 1983 \citep{Lasker:1990}, and three scans of photographic plates from the Palomar Observatory Sky Survey~II obtained in 1989-93 \citep{Reid:1991}. Data analysis was performed using our own astrometric software with the {\it Gaia} DR3 catalogue as reference. To reduce the impact of the large time difference between the {\it Gaia} catalogue and the observation epoch, the reference stars were selected in an iterative process. Only stars fitting within 0.1 arcsecs in the third-order astrometric plate solution were kept and used in the final solution.  The internal statistical uncertainties of the positions obtained were estimated using the bootstrap sampling procedure \citep{Efron:1982}. These measurements are presented in Table \ref{tab:position2} together with a collection of catalogue positions of HD\,7977 we were able to find, all transformed to ICRS.

Listed catalogue positions are often not independent but are based on a different combination of observations. If a mean epoch and a proper motion were included in a catalogue, we propagated the star's position to this epoch using the given proper motion. A rigorous transformation to the ICRS system has also been performed.

It should be stressed that in almost all cases of historical catalogues, as uncertainty, we use the formal precision of the published position (typically 0\fs01 and 0\farcs1). To our knowledge, this underestimates the corresponding probable errors by a factor of three to five or more. For example, in the "Catalog von 14680 Sternen" \citep{Krueger:1890} an author gives an estimate of the probable error of the single observation equal to 0\fs101 and 0\farcs51 for Right Ascension and Declination, respectively. But these values are presented in this catalogue with a precision of 0\fs01 and 0\farcs1, respectively.

\begin{table*}
        \caption{ICRS positions of HD\,7977 from various catalogues and our own measurements based on several photographic surveys (see \ref{sec:astrometry-our} for details). 
        The first two columns present the Right Ascension and Declination epochs; the next four columns contain positions and their uncertainties. The eighth column gives references, and in the last column, the values we used in the proper motion estimation are marked with an asterisk. 
}
\label{tab:position2}
\begin{tabular}{l l l l l l l c c} 
        \hline
         &  &  & RA [sec] & Dec [arcsec] & RA error  & Dec error & & \\
        RA epoch & Dec epoch & Source & $1^h$ 20$^m$ & $+61$\degr 52\arcmin & [sec] & [arcsec] & Reference & Used \\
        \hline 
  1842.0    &  1842.0    & Argelander\tablefootmark{a}         &  31\fs69        &   57\farcs7        &   0\fs01       &     0\farcs1       & 1   &  * \\
  1875.0    &  1875.0    & Argelander\tablefootmark{a}         &  31\fs69        &   57\farcs8        &   0\fs01       &     0\farcs1       &  2  &  * \\
  1877.1    &  1877.1    & H+G Krueger\tablefootmark{a}        &  31\fs95        &   56\farcs6        &   0\fs01       &     0\farcs1       &  3  &   \\
  1903.94   &  1903.94   & VO\tablefootmark{a}                 &  31\fs676       &   57\farcs40       &   0\fs022      &     0\farcs195     &  4  &  * \\
  1903.94   &  1903.94   & our VO\tablefootmark{c}             &  31\fs627       &   57\farcs51       &   0\fs019      &     0\farcs17     &  this paper  &  * \\
  1903.94   &  1903.94   & our VO\tablefootmark{c}             &  31\fs675       &   57\farcs17       &   0\fs024      &     0\farcs15     &  this paper  &  * \\
  1929.8    &  1929.9    & AGK2\tablefootmark{a}               &  31\fs56        &   57\farcs3        &   0\fs01       &     0\farcs1       &  5  &  * \\
  1933.69   &  1932.62   & PPM\tablefootmark{a}                &  31\fs617       &   57\farcs18       &   0\fs016      &     0\farcs11      &  6  &    \\
  1943.816  &  1944.427  & ACRS\tablefootmark{a}               &  31\fs597       &   57\farcs25       &   0\fs022      &     0\farcs195     &  7  &  * \\
  1950.0    &  1950.0    & SAO\tablefootmark{a}                &  31\fs482       &   57\farcs64       &   0\fs024      &     0\farcs360     &  8  &   \\
  1951.81   &  1951.81   & AGK3U\tablefootmark{a}              &  31\fs586       &   57\farcs35       &   0\fs015      &     0\farcs106     &  9  &  * \\
  1954.8    &  1954.8    & our POSSI-E\tablefootmark{b}        &  31\fs5684      &   56\farcs863      &   0\fs0097     &     0\farcs043     & this paper   &  * \\
  1954.8    &  1954.8    & our POSSI-O\tablefootmark{b}        &  31\fs5530      &   56\farcs867      &   0\fs0051     &     0\farcs036     & this paper   &  * \\
  1983.123  &  1983.123  & FONAC\tablefootmark{a}              &  31\fs6095      &   57\farcs075      &   0\fs0047     &     0\farcs057     & 10   &  * \\
  1983.9    &  1983.9    & our Pal-QV\tablefootmark{b}         &  31\fs6270      &   56\farcs798      &   0\fs0038     &     0\farcs030     & this paper   &  * \\
  1989.68   &  1989.02   & UCAC4\tablefootmark{a}              &  31\fs6252      &   57\farcs061      &   0\fs0018     &     0\farcs027     &  11   &  * \\
  1989.8    &  1989.8    & our POSSII-J\tablefootmark{b}       &  31\fs5826      &   57\farcs169      &   0\fs0038     &     0\farcs016     &  this paper  &  * \\
  1991.00   &  1991.01   & PPMX\tablefootmark{a}               &  31\fs6539      &   57\farcs045      &   0\fs0011     &     0\farcs008     &  12  &   \\
  1991.10   &  1991.19   & Tycho-2\tablefootmark{a}            &  31\fs5956      &   57\farcs044      &   0\fs0011     &     0\farcs008     & 13   &  * \\
  1991.8    &  1991.8    & our POSSII-F\tablefootmark{b}       &  31\fs5805      &   57\farcs063      &   0\fs0027     &     0\farcs012     & this paper   &  * \\
  1993.8    &  1993.8    & our POSSII-N\tablefootmark{b}       &  31\fs5877      &   57\farcs127      &   0\fs0022     &     0\farcs008     &  this paper  &  * \\
  2000.0    &  2000.0    & ACT\tablefootmark{a}                &  31\fs5929      &   57\farcs010      &   0\fs0009     &     0\farcs006     &  14  &   \\
  2003.672  &  2003.672  & UCAC5\tablefootmark{a}              &  31\fs61012     &   57\farcs0661     &   0\fs00003    &     0\farcs0003    &  15   &  * \\
  2013.550  &  2013.550  & URAT1\tablefootmark{a}              &  31\fs5924      &   57\farcs189      &   0\fs0016     &     0\farcs011     & 16   &  * \\
  2015.0    &  2015.0    & \textit{Gaia} DR1\tablefootmark{a}  &  31\fs596488    &   57\farcs07803    &   0\fs000031   &     0\farcs00030   &  17  &   \\
  2015.5    &  2015.5    & \textit{Gaia} DR2\tablefootmark{a}  &  31\fs5965293   &   57\farcs0780133  &   0\fs0000032  &     0\farcs0000261 &  18  &   \\
  2016.0    &  2016.0    & \textit{Gaia} DR3\tablefootmark{a}  &  31\fs5965296   &   57\farcs0779238  &   0\fs0000028  &     0\farcs0000236 &  19  &  * \\
    
    \hline
\end{tabular}
\\
        \tablefoottext{a}{star catalogue} \hspace{1cm}
        \tablefoottext{b}{scanned photographic plate astrometry} \hspace{1cm}
        \tablefoottext{c}{measured photographic plate astrometry}

        \tablebib{
         (1) \cite{Oeltzen:1842}; (2) ; (3) \cite{Krueger:1890};    (4) \cite{Urban-Vatican-Vizier:1996}; (5) \cite{AGK2-1951}; 
         (6) \cite{ppm_cat:1988}; (7) \cite{ACRS:1991}; (8) \cite{SAO:1966}; (9) \cite{AGK3-1975}; (10) \cite{FONAC:2016}; (11) \cite{zacharias-UCAC4:2012}; (12) \cite{PPMX:2008}; (13) \cite{tycho2-cat:2000}; (14) \cite{ACT:1998}; (15) \cite{UCAC5:2017}; (16) \cite{URAT1:2015}; (17) \cite{GaiaDR1:2016}; (18) \cite{Gaia-DR2:2018}; (19) \cite{Gaia-DR3-release:2022};
        
        }
\end{table*}        

\begin{figure}
    \centering
    \includegraphics[width=0.35 \textheight]{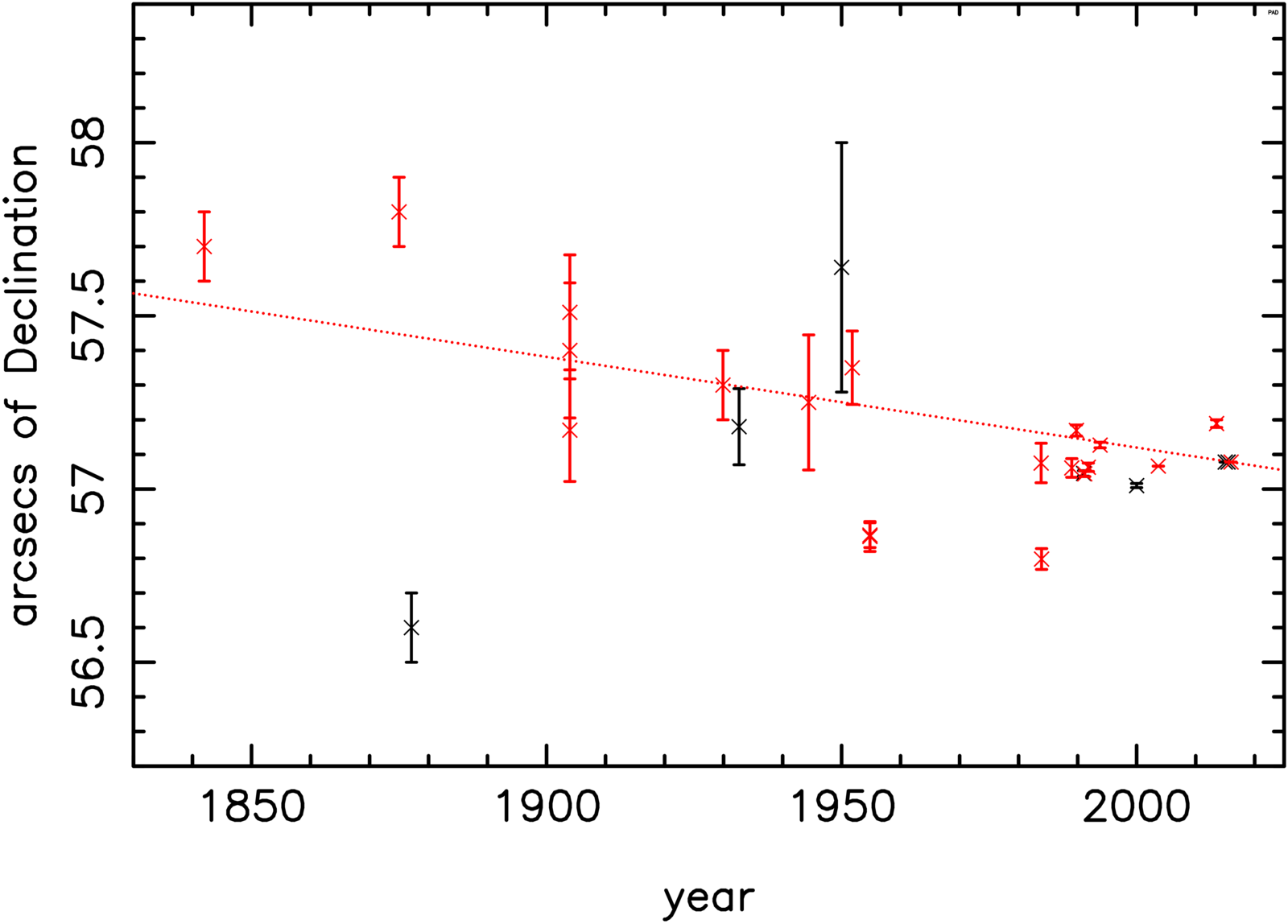}
   \includegraphics[width=0.35 \textheight]{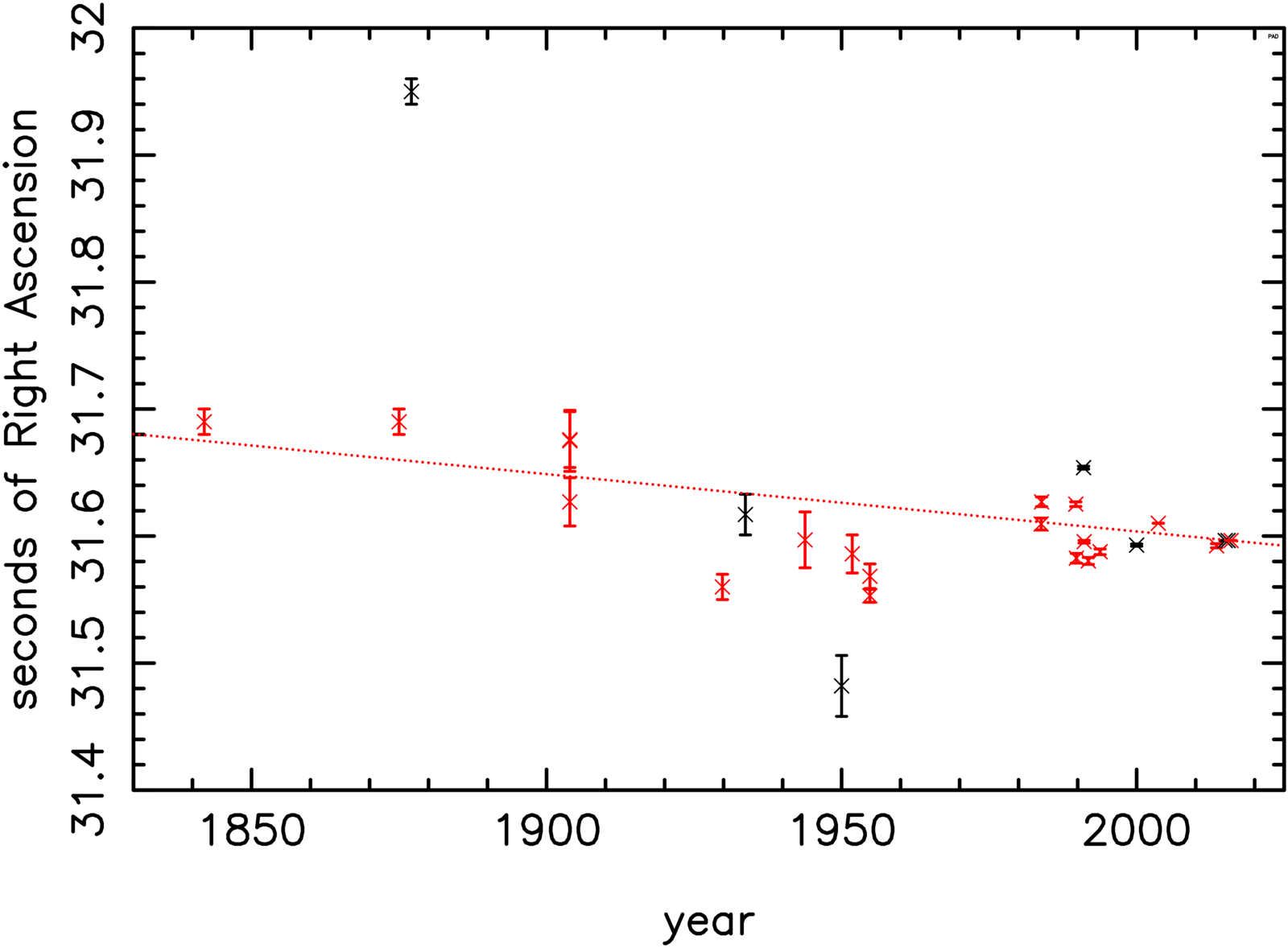}
    \caption{HD\,7977 position changes over last two centuries. Data are from Table \ref{tab:position2}, in red we show these which were taken into account in proper motion estimation. No weighing has been applied but straight lines are forced to pass through the {\it Gaia} DR3 position. The estimated proper motion components are $-2.62$ mas/yr and $-3.19$ mas/yr in Declination and Right Ascension respectively.}
    \label{fig:positions1}
\end{figure}

\subsection{Our attempt to estimate HD\,7977 proper motion}

Using data from Table \ref{tab:position2}, we attempted to determine the proper motion of this star. We assumed a linear trend in both coordinates, as no information on HD\,7977 duplicity can be found in any source. Moreover, in some catalogues, the authors comment that, after some checks, the possibility of its double nature has been rejected.

The results of our estimation are presented in Fig.~\ref{fig:positions1}. Some of the positions have been excluded from our straight-line fitting. We exclude the data obtained by \cite{Krueger:1890} as an evident outlier. We also exclude SAO, PPM, and PPMX positions since these catalogues are compilations of different data, as is the case for the ACT catalogue. We also exclude {\it Gaia} DR1 and DR2 positions by assuming that the DR3 data are the most accurate and taking into account a small time interval between them. The excluded data are presented in black in Fig.~\ref{fig:positions1}. 

When fitting the straight line, we forced it to pass through the {\it Gaia} position assuming that even in future data releases the position would change infinitesimally on the scale of Fig.~\ref{fig:positions1}. It should be stressed again that the error bars of all positions before 1940 are underestimated and should be multiplied by 3 or 5. On the other hand, in the case of such a small proper motion, all historical observations are of great value.

From the linear fit presented in Fig.~\ref{fig:positions1} we obtained the following HD\,7977 proper motion components $\mu_{\alpha}\cdot \cos(\delta) = -1.23$ mas/yr and $\mu_{\delta} = -3.19$ mas/yr. We do not attempt to estimate their uncertainties due to the mentioned fact that the probable errors of the historical positions are unknown. The value obtained seems almost consistent with the proper motions of Tycho-2 and PPMX.

It is worth mentioning that such a small proper motion might be very difficult to measure. For example, the Vatican Observatory plates from 1903 were measured manually with a formal precision of 0\farcs3. Even if we ignore any (quite probable) random errors and try to estimate proper motion comparing the Vatican observations with the {\it Gaia} data, we obtain the formal uncertainty of 0\farcs3$ / 120 yr =$ 2.5 mas/yr which is comparable to a proper motion value of HD\,7977. If we account for random observational errors in historical data, we may conclude that a very small proper motion of this star presented in {\it Gaia}~DR3 can be consistent with all previous observations. As a result, we have to accept the possibility of a very close passage of HD\,7977 near the Sun.

\subsection{Radial velocity}\label{sec:radial-velocities}

The first measurement of HD\,7977 radial velocity was published as a part of {\it Gaia}~DR2 catalogue \citep{Gaia-DR2:2018}. As a result, the past motion of this star could be analysed and its moderately close pass near the Sun was first announced by \cite{Bailer-Jones:2018}. Using the same data, \cite{First-stars:2020} presented the possible interaction of this star with the comet C/2002~A3~LINEAR. The second radial velocity measurement for HD\,7977 was published by \cite{Errmann:2020} owing to a dedicated campaign for stars that closely passed the Sun. Using this later value combined with the astrometry from {\it Gaia}~EDR3, \cite{Dyb-Ber-StePPeD:2022} announced an extremely close nominal passage of HD\,7977 near the Sun 2.5\,Myr ago at a distance of 0.014\,pc (~3,000\,au).

The next value was included in the third {\it Gaia} data release \citep{Gaia-DR3-release:2022} and was used by \cite{Bob-Baj:2022} and \cite{Bailer-Jones:2022} to estimate the parameters of this star's close passage near the Sun. The comparison of all these velocity measurements is presented in Table \ref{rad_vel}.

\begin{table}
        \caption{Radial velocity measurements for HD\,7977.}
        \label{rad_vel}
        \begin{tabular}{l c c } 
        \hline
        Source & Radial Velocity & RV error         \\
               & [km s$^{-1}$]   & [km s$^{-1}$]    \\
        \hline  
        {\it Gaia} DR2 (2018) & 26.45 & 0.35   \\
        Errmann et al. (2020) & 29.93 & 0.28   \\
        {\it Gaia} DR3 (2022) & 26.76 & 0.21   \\
        \hline
        \end{tabular}
\end{table}

We  checked most of the contemporary stellar databases looking for other determinations of radial velocity and other stellar parameters of the investigated star. According to available data, there are no planets orbiting HD\,7977 and no stellar companion has been found so far\footnote{https://exofop.ipac.caltech.edu/tess/target.php?id=444593991}. In light of the discussion of the quality of {\it Gaia} radial velocity measurements presented in Sect.~\ref{sec:comment_on_gaia}, we decided to use the value published by \cite{Errmann:2020} in our calculations.

\section{The influence of the data uncertainties on the HD\,7977 close passage parameters}\label{sec:HD7977_uncertainties_vs_close_pass}

Taking the astrometry from  {\it Gaia}~DR3 and the radial velocity from \cite{Errmann:2020}, we obtained the nominal distance between HD\,7977 and the Sun to be equal 0.011\,pc ($\sim2,300$\,au) during its passage 2.47\,Myr ago. This is an even slightly closer passage than that obtained by \cite{Dyb-Ber-StePPeD:2022} based on {\it Gaia} EDR3. The difference is the result of the modified list of other acting stars, which comes from partially new radial velocities in  {\it Gaia}~DR3.

\begin{figure}
    \centering
    \includegraphics[width=0.35 \textheight]{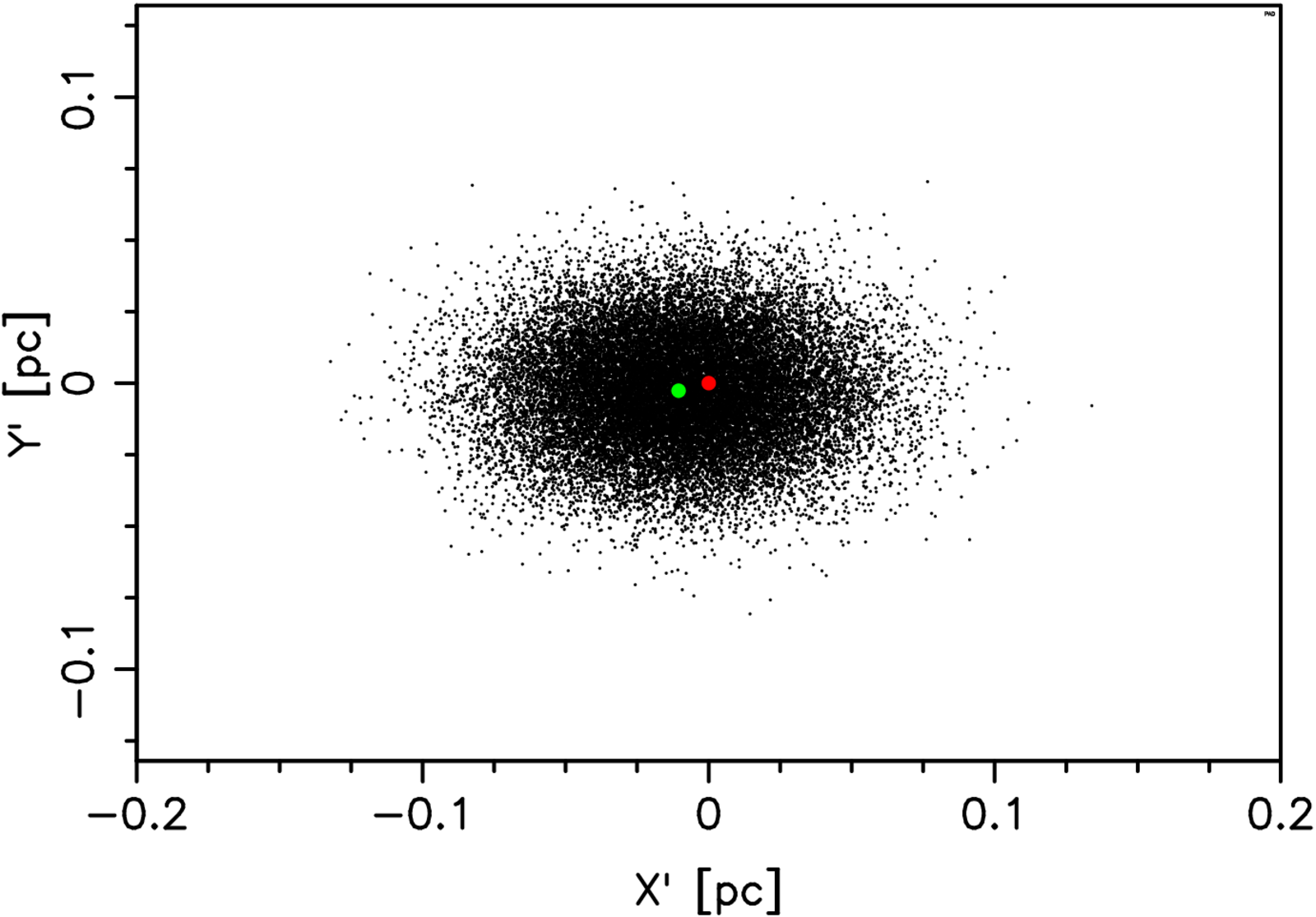}
    \caption{HD\,7977 and its 25,000 clones at the closest encounter with the Sun. The red dot represents the Sun position while the green one corresponds to the star's nominal position at the closest approach. }
    \label{fig:P0230_scatter}
\end{figure}

To estimate the uncertainty of these parameters resulting from the uncertainty of the HD\,7977 data, we cloned this star 25\,000 times using the covariance matrix from {\it Gaia} DR3. Next, we numerically integrated all clones into the past and calculated the parameters of their closest encounter with the Sun. The result of this simulation is shown in Fig.~\ref{fig:P0230_scatter}.  The details of such a calculation including the discussion on the Galactic gravitational potential models can be found in \cite{Dyb-Ber-StePPeD:2022}. 

The formal statistics of the minimal distance results for 25,000 star clones resulted in $d_{min} = 0.012 : 0.031 : 0.060$ pc, which is presented here in the form of the $10^{th}$ percentile, the median, and the $90^{th}$ percentile.  Analogous statistics for the moment of the closest approach read: $t_{min} = 2.44 : 2.47 : 2.50$\,Myr ago and for the relative velocity of the encounter: $v_{rel} = 30.3 : 30.6 : 31.0$\,km$\cdot$s$^{-1}$.

The cloud of star clones presented in Fig.~\ref{fig:P0230_scatter} surrounds the Sun, so the formal statistics of the minimal distance are less informative. Instead, we calculated the distance between the Sun and the centroid of this cloud to be equal to 0.011\,pc. We can also state that $90\%$ of the clones lie no further than 0.058\,pc from this centroid. The main concern with this stellar passage is that it was very close to the Sun but its precise trajectory cannot be reliably calculated because of the still large astrometry uncertainties.

\begin{figure}
    \centering
    \includegraphics[width=0.35 \textheight]{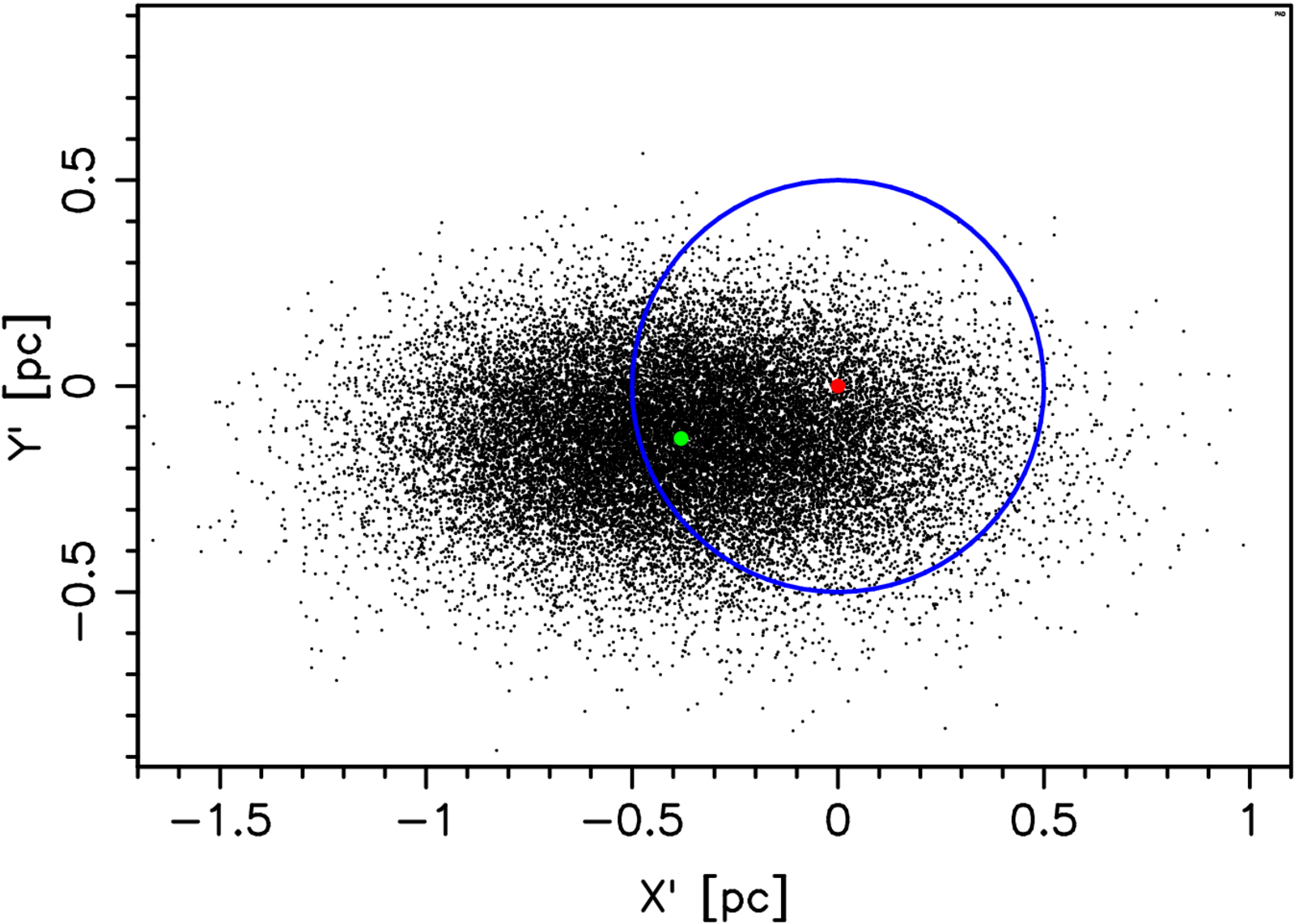}
    \caption{P3509 and its 25,000 clones at the closest encounter with the Sun. The red dot represents the Sun position while the green one corresponds to the star's nominal position at the closest approach. The blue circle marks the traditional extent of the Oort cometary cloud. }
    \label{fig:P3509_scatter}
\end{figure}

\section{Other important stellar perturbers -- the StePPeD database update}
\label{sect-other-perturbers}

Based on our current knowledge the nominal passage of HD\,7977 near the Sun is the closest one, but we have recognised tens of other potential stellar perturbers of LPCs motion. In July 2022 {\it Gaia} DR3 was released \citep{Gaia-DR3-release:2022}. No new astrometry for single stars has appeared, but many corrected and new radial velocities have become available. The catalogue contains the radial velocities for about 33 million stars, which is 5 times more than in Gaia EDR3. Moreover, for the first time in the {\it Gaia} catalogue there appeared a special part with the new astrometric results for over 800,000 unresolved binaries with their observations treated with a new, dedicated model \citep{gaia-dr3-binaries:2022}.

With all these new data at hand, we decided to update the StePPeD database from version 3.2 described in \citet{Dyb-Ber-StePPeD:2022} to a new one, namely release 3.3, which was made publicly available in June 2023. According to our own cometary calculations experience and following a similar approach of \cite{Bob-Baj:2022} and \cite{Bailer-Jones:2022} we decided to limit the list of potential perturbers to stars that had passed or will pass the Sun closer than 1\,pc. There are 59 such stars in our new list. All parameters of known close encounters with the Sun have been recalculated using the corrected radial velocities. Additionally, we searched for new stars using new radial velocities. We have found 26 new perturbers with a miss distance smaller than 1\,pc.

We also decided to check all stars suspected to be unresolved binaries and included in the {\it Gaia} DR3 Part 3, non-single stars. We found that for two stars from our list, namely P0180 and P0533, we should use the astrometry from tables describing nonsingle-star astrometric models for sources having a nonlinear proper motion which is compatible with an acceleration solution, acceleration model with 9 parameters (for P0180) and with 7 parameters (for P0533). These new data improved our results, for example, the closest approach for P0533 decreased from 0.51$\pm$0.12\,pc down to 0.23$\pm$0.05\,pc. But the most important new result was the discovery of a new massive perturber encoded in the StePPed database as P3509.

\subsection{The special case of P3509}
\label{subsect-P3509}

During our careful inspection of all results presented in {\it Gaia} DR3 for unresolved nonsingle stars, we found that star {\it Gaia} DR3 3598849446322133248, also known as \object{HD 102928} or HIP\,57791, is a binary system that nominally approached the Sun 3.94\,Myr ago at a distance of only 0.40\,pc. In the StePPeD database, it is named P3509 and we will use this in the following text.

The uncertainties in astrometry and radial velocity of P3509 are significantly higher than for HD\,7977 and result in a much more spread clones swarm, as shown in Fig.~\ref{fig:P3509_scatter}. However, it clearly shows that the minimal distance between P3509 and the Sun can be arbitrarily small. What makes this potential perturber very important is its mass. According to \cite{Chulkov:2022} the total mass of this system equals 3.5\,M$_\odot$. All details on this particular system approach to the Sun, as well as on all other potential stellar perturber encounters, can be found in the StePPeD dabase. Basic data on stars that appear as perturbers in the examples presented in Section \ref{sec:LPCs_past_motion-examples} are also quoted in Table \ref{tab-stars-past} for the convenience of the reader.

\subsection{Multiple stars problem}

To our knowledge, up to 2022 there was no information in the literature about a recognised binary or multiple system that passed or will pass the Sun closer than 1\,pc. The only exception is \object{$\alpha$~Centauri} triple system which will approach the closest proximity of 1\,pc in the future (0.03\,Myr from now) but we have to wait for more precise data on this system to calculate precisely the parameters of this event. In \cite{Dyb-Ber-StePPeD:2022} we presented a list of double-star perturbers, but only two of them (P5000 and P5001) remain on our current shortlist. We should also note that there probably exist companions for P0109 (\object{HD 168769}, HIP\,90112) and P0506 from our current list, but no reliable data for secondaries are available, so we treated these perturbers as single stars.

In the course of preparing the next update of the StePPeD database, we used all new radial velocities from {\it Gaia} DR3 and analysed all putative binary systems included in \citet{million-binaries}. We found a large group of candidates for the close Solar System passage. Most of them have a very high uncertainty of the closest approach distance and its time, and until more precise data from {\it Gaia} are released (either in DR4 or DR5) these systems will remain only potentially interesting cases. Taking into account that a large percentage of stars are members of multiple systems \citep{2023ASPC..534..275O, 2023arXiv231109764M}, this line of research seems to be important.

\section{Possible effect of HD\,7977 and other stars on LPCs past motion}\label{sec:LPCs_past_motion}

The potentially extremely close passage of HD\,7977 near the Sun raises an obvious question about its influence on the motion of all bodies in the solar system. In the next sections, we discuss this subject in detail.

We start here with the influence of this stellar passage on the past motion of the long-period comets, since they are naturally the most sensitive bodies to external perturbations acting on their very elongated orbits. It is worth to note that stars can act on LPCs in two different ways:
\begin{itemize}
\item A star can pass very near a comet, especially when it is near its aphelion, tens of thousands au from the Sun. LPCs are Solar System bodies that reach the largest heliocentric distances.
\item A star can pass very near the Sun, thus perturbing solar motion in the Galaxy and, therefore, changing the orbits of all bodies of the Solar System at different scales. The most sensitive are again LPCs.
\end{itemize}

In the following sections, we present a simple statistical picture of the stellar perturbations acting on LPCs and several particular examples of the past evolution of cometary orbit under simultaneous Galactic and stellar perturbations. In both aspects, we focus on indicating the action of the star
HD\,7977.

\begin{figure}
    \centering
    \includegraphics[angle=90,width=0.35 \textheight]{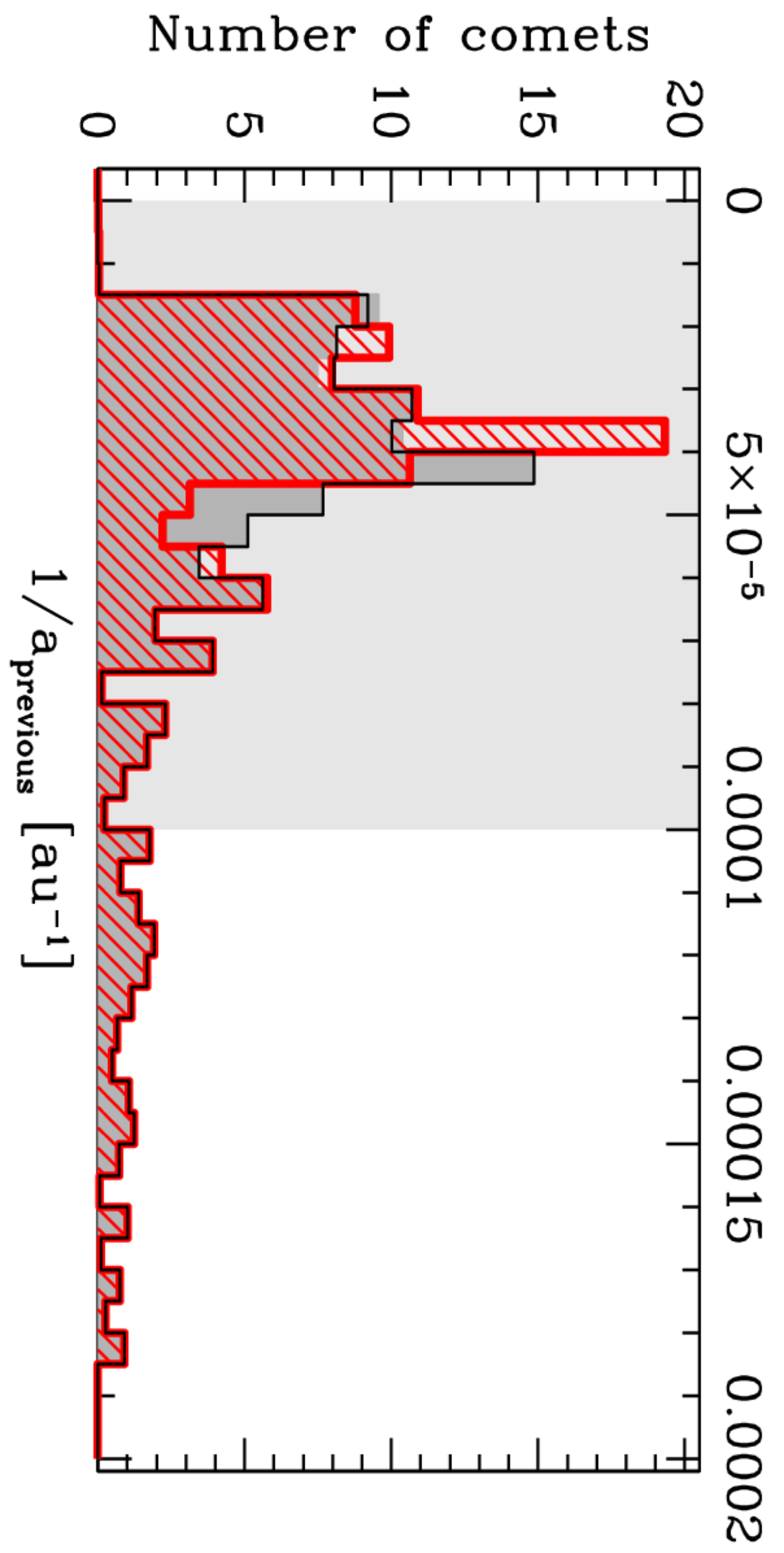}
    \includegraphics[angle=90,width=0.35 \textheight]{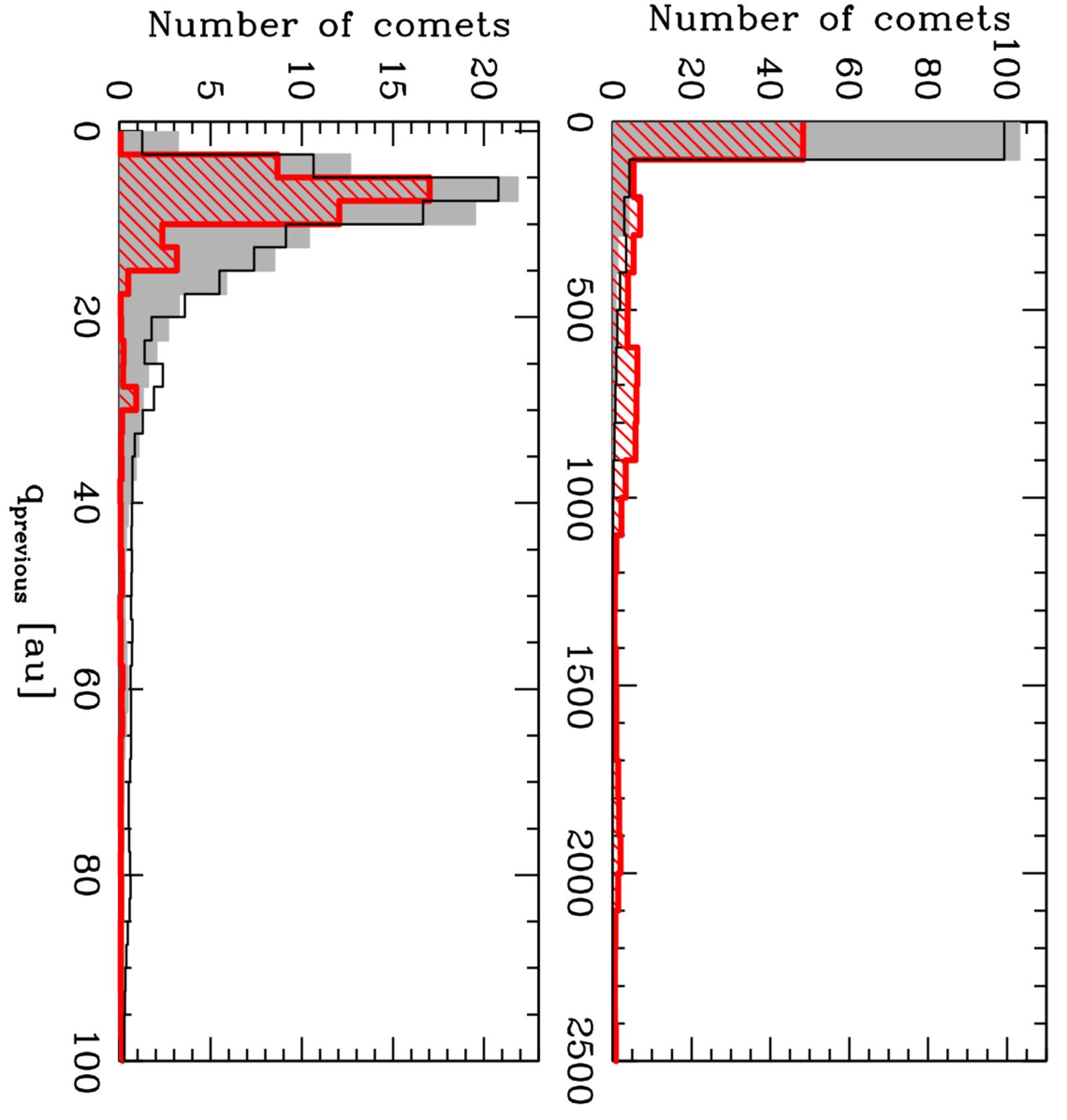}
    \caption{Statistics for 125~LPCs with the osculating perihelion distance greater than 3.5\,au described in Sect.~\ref{sec:comet-samples}. Upper panel: the distribution of previous $1/a$, where the filled gray histogram represents calculations with Galactic tide perturbations (stellar perturbations are ignored), the black-lined histogram represents results including perturbations from all stars except HD\,7977, and the red dashed histogram describes the case with perturbations from the Galactic tide and all stars. The light-grey vertical band shows the area traditionally called the Oort spike. The middle and lower panels show the distribution of the previous perihelion distance; the lower panel gives details about the first bin (1-100 au) of the middle panel.}
    \label{fig:Oort_spike_comets_hist}
\end{figure}

\begin{table*}
\caption{Star identifications and parameters of the close passages near the Sun for all eight objects that appear as perturbers in the examples presented in Section \ref{sec:LPCs_past_motion-examples}. D$_{min}$ and T$_{min}$ denote the nominal minimal Sun -- star distance and the epoch of the closest approach, respectively. The detailed information on this objects (including the uncertainties) can be obtained from the StePPeD database.} 
\label{tab-stars-past}
\centering
\begin{tabular}{c c c  c c c} 
\hline 
StePPeD ID & {\it Gaia} DR3 ID & Other names & D$_{min}$ [pc]  & T$_{min}$ [Myr] & mass estimate [M$_{\odot}$] \\
\hline
P0032 & 3228991219548947328 & \object{HD 32816}, HIP\,23714  & 0.806 & $-5.55$ & 2.01\\
P0230 & 510911618569239040 & HD\,7977, BD+61~250 & 0.011 & $-2.47$ & 1.08 \\
P0417 & 1281410781322153216 & \object{Ton 214}, WD\,1446+286 & 0.514 & $-1.47$ & 0.85 \\
P0506 & 5571232118090082816 & & 0.198 & $-1.08$ & 0.77 \\
P0514 & 6608946489396474752 & \object{2MASS J22415098-2759470} & 0.577 & $-2.78$  & 0.75  \\
P0533 & 3118526069444386944 & & 0.223 & $-3.19$ & 0.87 \\
P3006 & 2773250629157880192 & & 0.231 & $-6.90$ & 0.42 \\
P3509 & 3598849446322133248 & HD\,102928, HIP\,57791 & 0.402 & $-3.94$ & 3.50 \\
\hline
\end{tabular}
\end{table*}

\subsection{The sample of investigated LPCs}\label{sec:comet-samples}

To study the influence of HD\,7977 and other stars on comets discovered to date, we selected a subset of the original orbital solutions of 312 long-period comets from the CODE catalogue\footnote{\url{https://pad2.astro.amu.edu.pl/CODE}} \citep{kroli-dyb:2020,Kroli-Dyb:2023}.  This source contains an almost complete sample of comets with original semimajor axes greater than 10,000\,au, and discovered in the period 1901--2021~January.

To minimise the impact of nongravitational effects (NG) on the motion of the comets discussed, we selected comets with perihelion distances exceeding 3.5\,au for further studies of their evolution in the past. In most cases of well-observed comets with a large number of positional observations, we use the solutions dedicated to the studies of the past dynamics and based only on the preperihelion observations. This method has recently been discussed by \citet{Kroli-Dones:2023}. As a consequence of this approach, we also exclude from our sample all comets discovered after their perihelion passage.

Finally, we conducted statistical investigations on 125 comets. In this sample, we have only comets with orbits of very good accuracy, classes 1a+, 1a, and 1b in the classification of \cite{kroli-dyb:2013}. It is worth noting that finally in this sample there are only 9~comets discovered before 1980, and as many as 86 discovered after 2000.

\subsection{The statistical picture of the HD\,7977 influence on comets}\label{sec:LPCs_past_motion-statistics}

The overall influence of stars on the past evolution of Oort Cloud comets is important. To our knowledge, the dominant star is HD\,7977, but the influence of other stars cannot be ignored. We demonstrate this in Fig.~\ref{fig:Oort_spike_comets_hist} using a sample of 125 comets with the observed osculating perihelion distance greater than 3.5\,au (see the previous section for an explanation). Typically, the stellar perturbations do not change the comet's semimajor axis much, but the overall distribution of the previous semimajor axes reveals some differences, as shown in the upper part of Fig.~\ref{fig:Oort_spike_comets_hist}. The uncertainties of comet orbits are taken into account in this figure by using an additional 5,000 orbit clones; for more details on how we construct such distributions, see, for example, \citet{kroli_dyb:2017}. The bin width for 1/a was assumed 0.000005\,au$^{-1}$, so it could be compared with Fig.~12 in \cite{krolikowska:2020}.

The filled grey histogram of $1/a_{\rm previous}$ represents the evolution taking into account only perturbations related to the Galaxy. The position of the maximum distribution in this case is between 40 and 45 in units of $10^{-6}$\,au$^{-1}$. The black-lined histogram indicates the result of the calculations that include all stars from the latest StePPeD database release except for HD\,7977, and is almost the same as the grey histogram. However, after including HD\,7977, we obtain a different distribution with the maximum of $1/a_{\rm previous}$ shifted towards less tight orbits, to the value between 35 and 40 in units of $10^{-6}$\,au$^{-1}$ (dashed-red histogram).  

The next two panels presented in Fig.~\ref{fig:Oort_spike_comets_hist} show how notable stars change the distribution of previous perihelion distances in the range between 0 and 2,500\,au (middle panel) and in the range of 1 to 100\,au (lowest panel). From the first bin in the middle panel, one can calculate that only 36.3\% of the comets in our sample had a previous perihelion distance below 100\,au when perturbations from all stars are included in past dynamical evolution. Moreover, our calculations indicate the dominant role of the star HD\,7977 here, which is illustrated by similar numbers (percent) of comets in the first bin for calculations without stellar perturbers (filled grey histogram, 75.5\% of all orbits) and including perturbations from stars except HD\,7977 (black-lined histogram, 74.3\%). In other words, HD\,7977 is solely responsible for about 40\% of previous perihelion distances greater than 100\,au, in many cases even significantly greater, while perturbations by Galaxy tides are responsible only for about 20\%. Detailed statistical research of this sample of comets will be the subject of a separate publication. Some examples of the dominant role of HD\,7977 in previous calculations of the perihelion distance are given in the next section.

\subsection{Selected examples of a strong interaction of HD\,7977 with comets} \label{sec:LPCs_past_motion-examples}

\begin{figure*}
    \centering
    \includegraphics[width=0.35 \textheight]{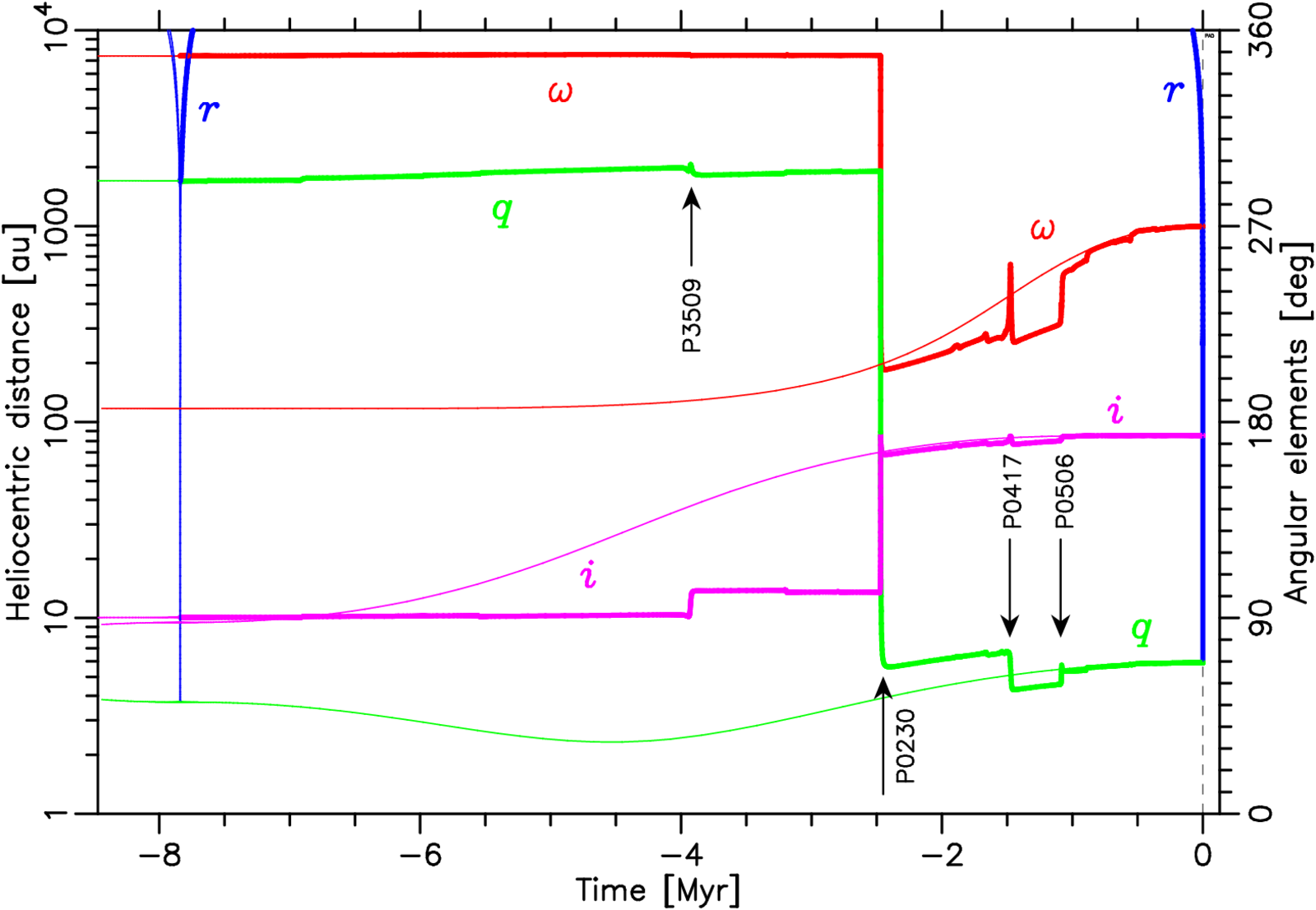}
    \hspace{0.4cm}
    \includegraphics[width=0.35 \textheight]{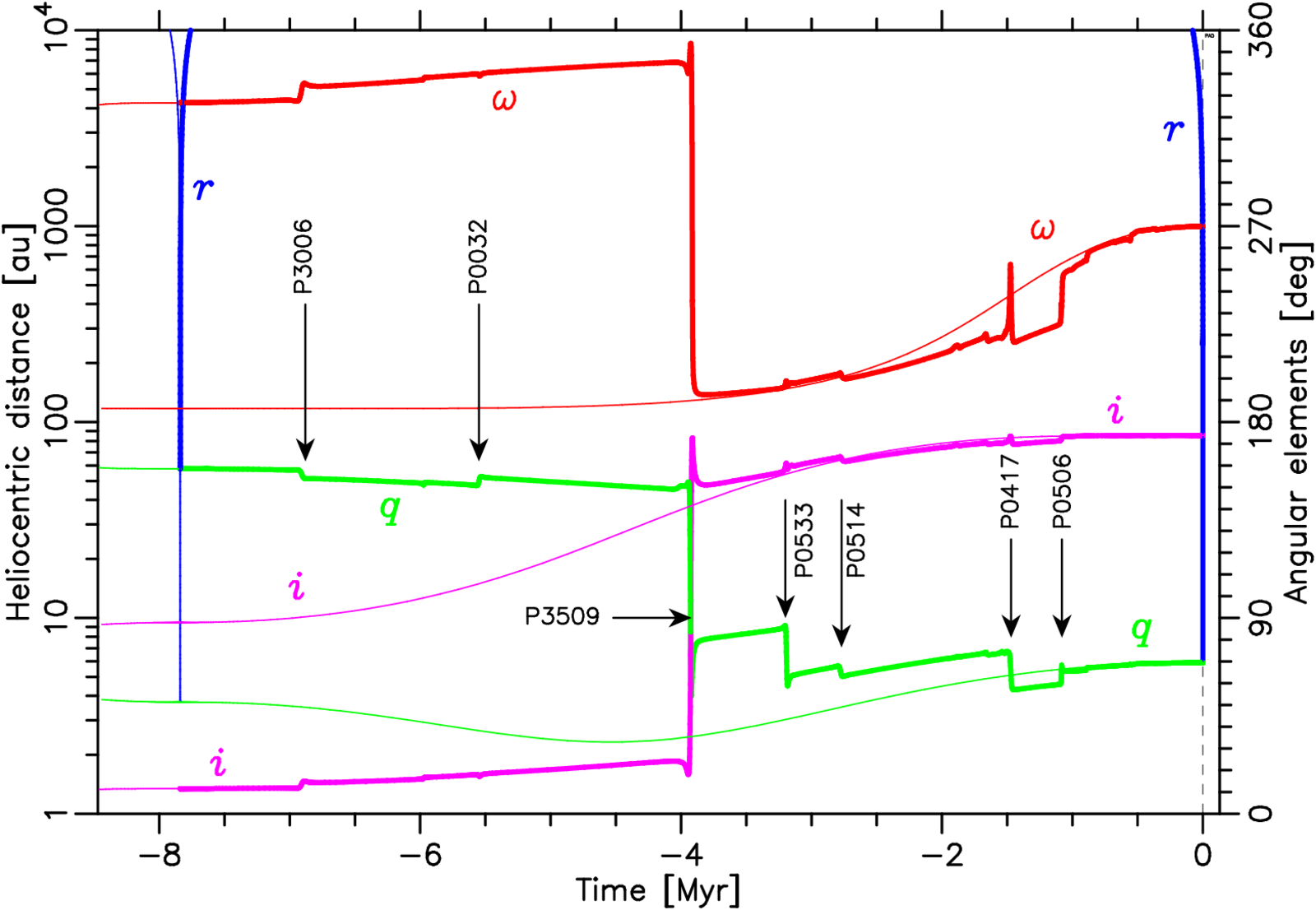}
    \caption{Past dynamical evolution of C/2010~S1 nominal orbit (“pa” solution, see the CODE Catalogue). Changes in the perihelion distance (green), inclination (fuchsia), and the argument of perihelion (red) are shown; angular elements are expressed in the Galactic frame. The thick lines show the result of the full dynamical model, while the thin lines show the evolution of elements in the absence of any stellar perturbations. The picture on the left -- all stars included, the picture on the right -- all stars included, except HD\,7977 (P0230 in the plot). In both pictures, noticeable changes in the perihelion distance are marked with arrows accompanied by the name of the corresponding stellar perturber (see Tab.\ref{tab-stars-past}).}
\label{fig:2010s1_past_evolution}
\end{figure*}

\begin{figure*}
    \centering
    \includegraphics[width=0.35 \textheight]{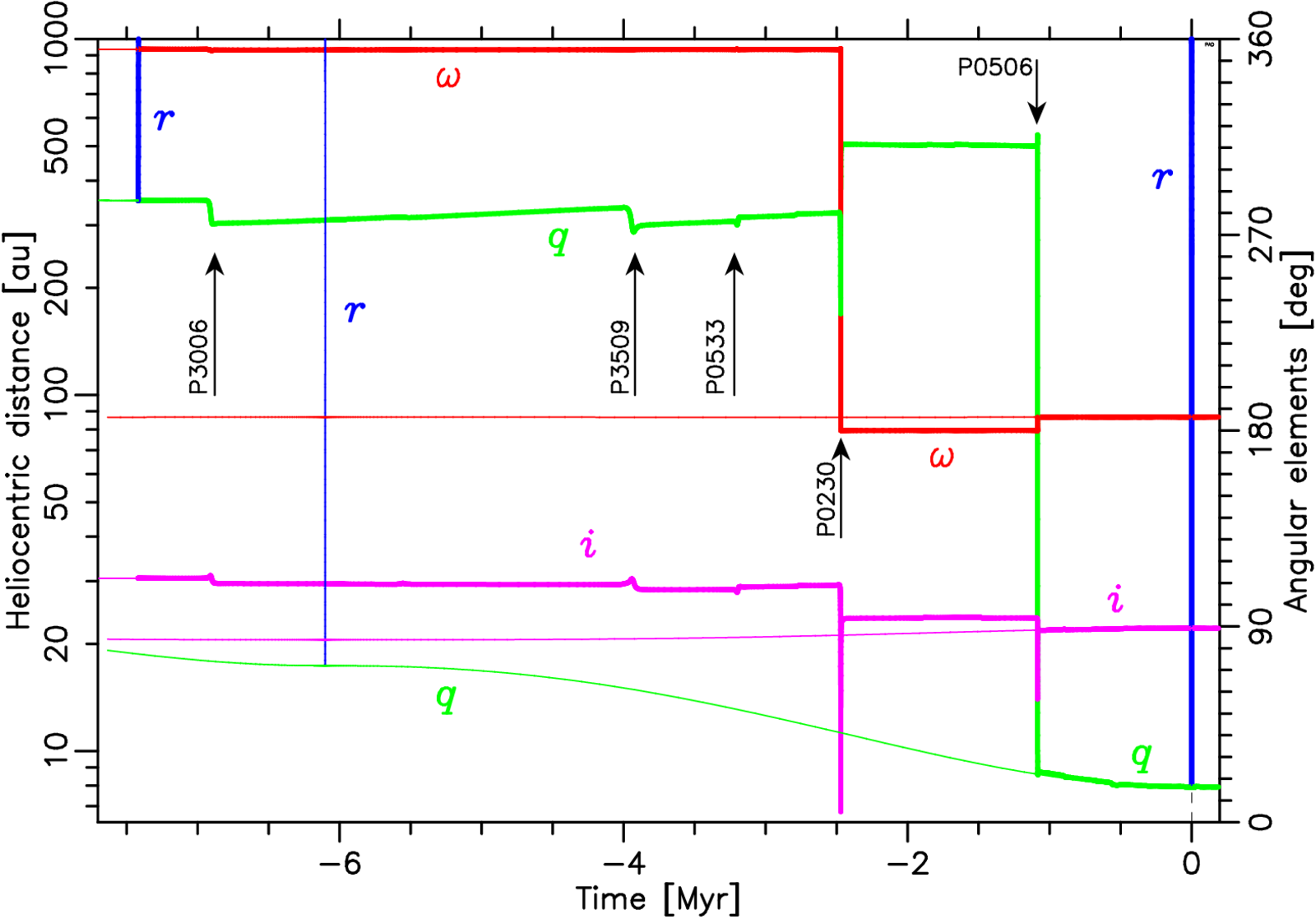}
    \hspace{0.4cm}\includegraphics[width=0.35 \textheight]{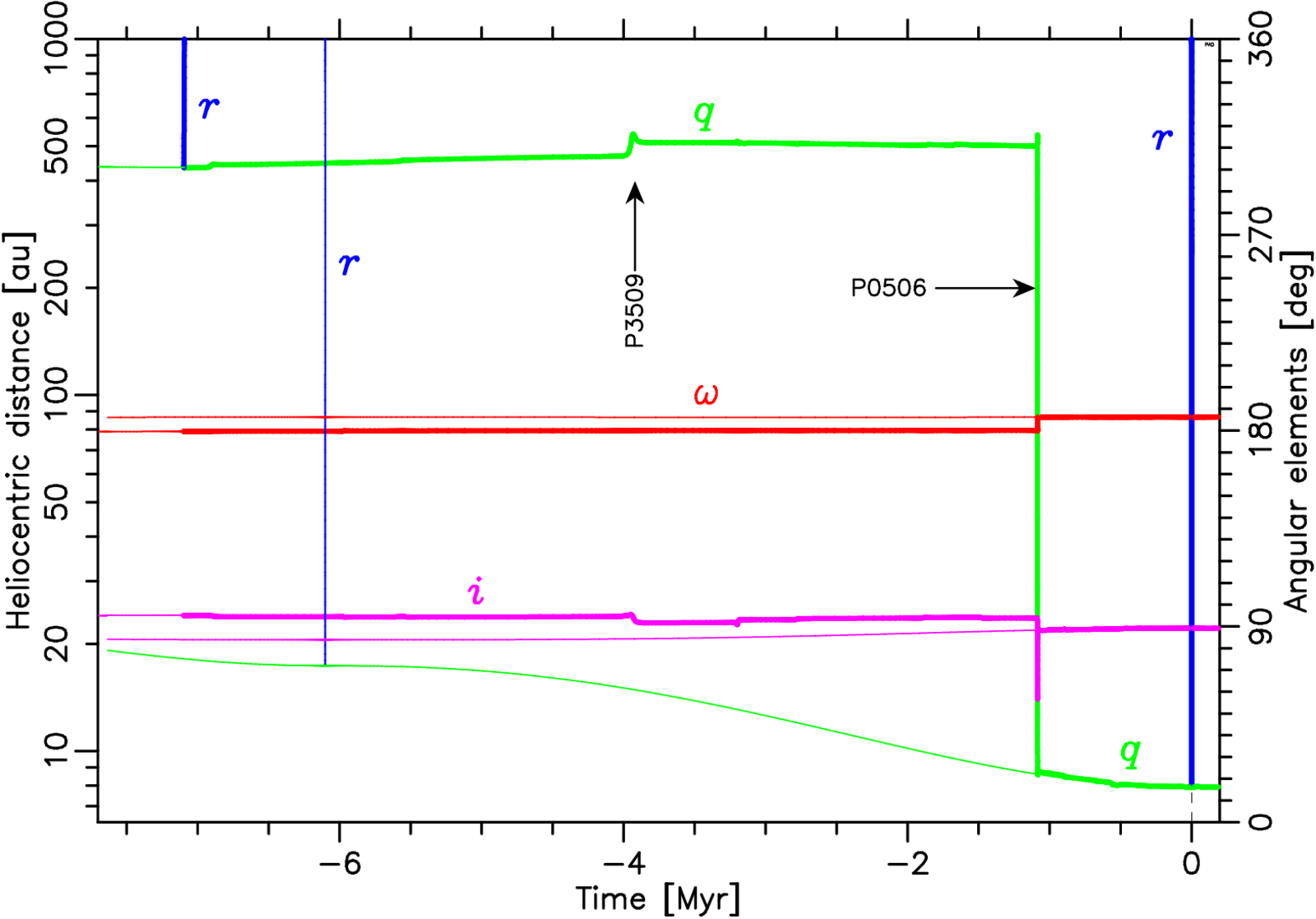}
    \caption{Past dynamical evolution of C/2015~XY$_1$ nominal orbit (“ba” solution, see the CODE Catalogue). Colour-coded curves are described in the previous pictures. Left picture -- all stars included, right picture -- only HD\,7977 excluded. Several stellar perturbations are marked.}
\label{fig:2015xy1_past_evolution}
\end{figure*}

As we have already mentioned, HD\,7977 (P0230) has passed close to the Sun 2.47\,Myr ago. Other stars with potentially strong influence are P0506 with the closest approach 1.08\,Myr ago to a distance of 0.2 pc and P0417 with a close encounter at 1.47\,Myr ago at 0.5\,pc. The perturbations from these two mentioned stars are visible in the figures presented below, but generally, they are weaker than the influence of HD\,7977. However, there are comets where the influence of other stellar perturbers is stronger than HD\,7977, which comes from the geometry and time coincidence of the star and comet orbital motion, see for example the past dynamics of C/2015\,XY$_1$. 

Below we show different stellar action examples on selected comets in order from their longer to shorter original semimajor axes, that is, from their orbital periods longer to shorter. All presented cases are typical and representative. 

The first three comets have previous orbital periods of almost 8\,Myr (C/2010~S1), $\sim$6\,Myr (C/2015~XY$_1$) and $\sim$4\,Myr (C/2012~LP$_{26}$), respectively. The fourth case of C/2010\,U3 represents a special group of comets that have a previous orbital period shorter than 2.47\,Myr, so they cannot meet HD\,7977  during their last revolution. In most of such cases, the previous perihelion distance is within the planetary zone. We cannot calculate the path of a comet through it due to uncertainties of the planetary position, so there is no reliable possibility to study the influence of HD\,7977 on such comets further in the past. However,  the C/2010\,U3 shows that the previous perihelion distance can be large.

\subsubsection{C/2010~S1 (LINEAR)}

The dynamical evolution of C/2010~S1 shows a rather complicated case of multiple potential stellar perturbations acting on this comet orbit. Fig.~\ref{fig:2010s1_past_evolution} shows the past evolution of C/2010~S1 using purely gravitational orbit (GR) based on preperihelion data ('pa' in the CODE catalogue, the highest quality class of 1a+). This comet passed its previous perihelion 7.65\,Myr ago. Taking into account only Galactic perturbation, we find that the previous perihelion distance, $q_{\rm prev}$, was 3.45~--~3.72~--~4.02~au (the 10$^{th}$ percentile, the median and the 90$^{th}$ percentile, where only uncertainties of a comet's orbit are taken into account). Including perturbations of all stars and going back in time, we notice that HD\,7977 was responsible for a large change in this comet's osculating perihelion distance, from almost 2,000\,au to less than 10\,au. This means that, according to our calculations, this comet was nominally moving on orbit with a heliocentric perihelion distance of 1691~-~1704~-~1716~au (same three percentiles) before being perturbed by this star. 

\begin{figure*}
    \centering
    \includegraphics[width=0.35 \textheight]{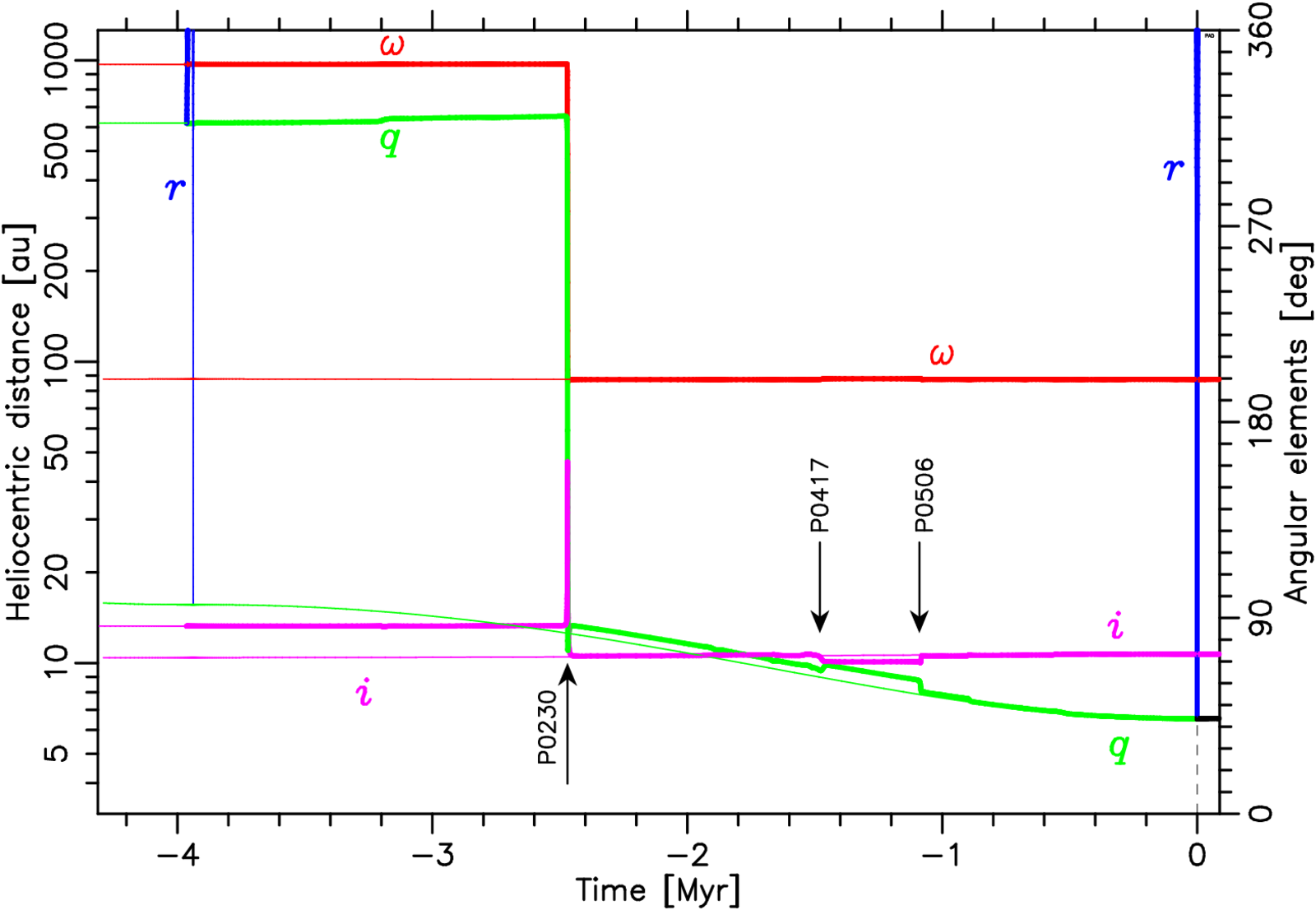}
    \hspace{0.4cm}\includegraphics[width=0.35 \textheight]{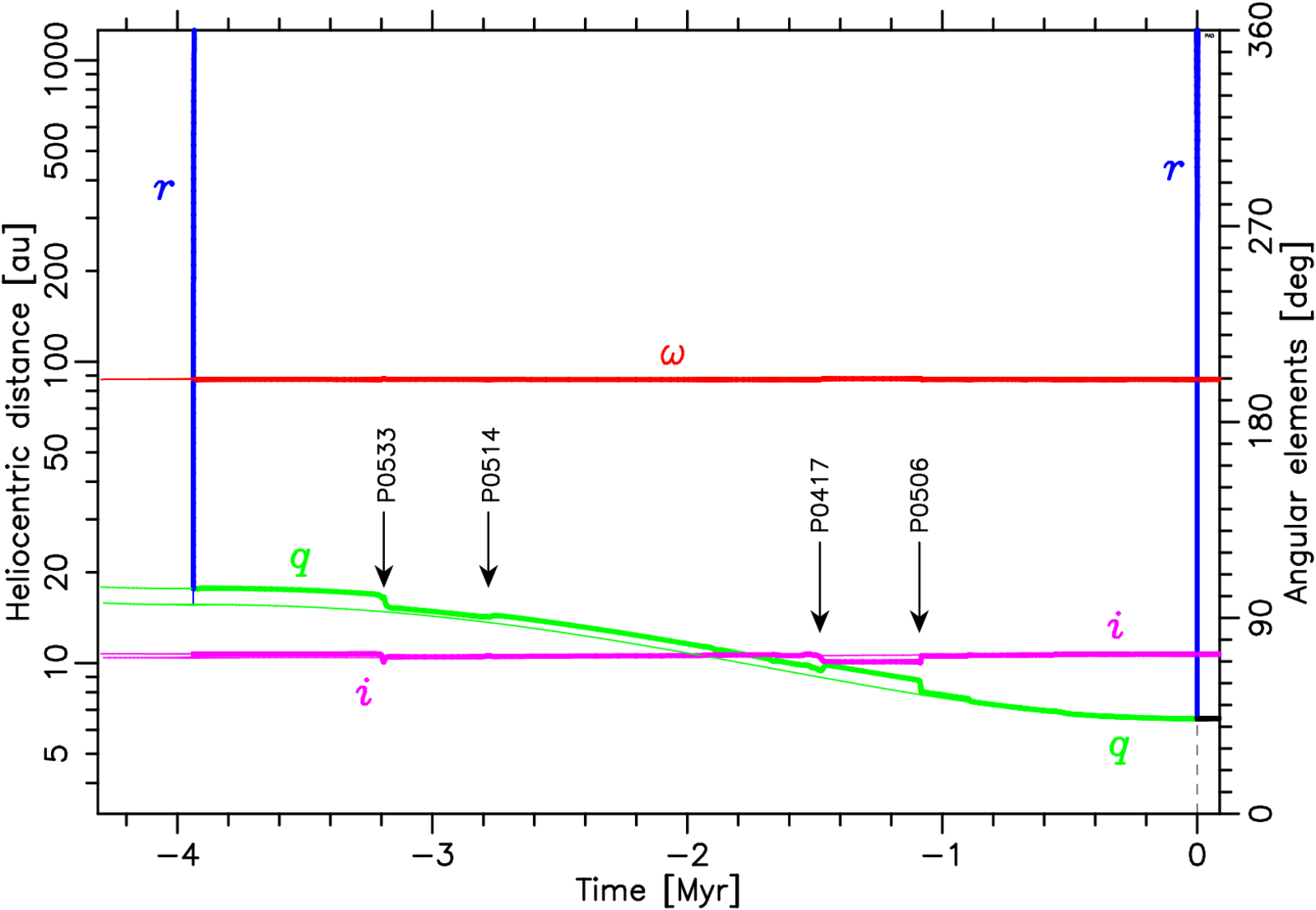}
    \caption{Past dynamical evolution of C/2012~LP$_{26}$ nominal orbit (“aa” solution, see the CODE Catalogue). Colour-coded curves are described in the previous pictures. Left picture -- all stars included, right picture -- only HD\,7977 excluded.}
    \label{fig:2012lp26_past_future_evolution}
\end{figure*}

During our backward numerical integration, we also noticed small perturbations from stars P0506 and P0417. These names come from the StePPeD database and, for the convenience of the reader, in Table \ref{tab-stars-past} we present their conventional names and nominal parameters of their close passages near the Sun.

The right panel of Fig.~\ref{fig:2010s1_past_evolution} shows the orbital evolution of C / 2010\,S1 when we exclude HD\,7977 from the dynamical model.  Then, we can notice much smaller stellar perturbations and a rather strong action of a massive perturber P3509. Note that this star does not act noticeably when HD\,7977 is in action (see left part of Fig.~\ref{fig:2010s1_past_evolution}).

\subsubsection{C/2015~XY$_1$ (Lemmon)}

The dynamical evolution of C/2015~XY$_1$ displays a very interesting case of a strong double stellar action in the past, namely by P0230 (HD\,7977) and P0506. This comet passed the current perihelion in April 2018 and was observed for four years at heliocentric distances between 9.52 -- 7.93 (perihelion) -- 9.62~au. Taking into account only galactic perturbation, we find that the previous perihelion distance, $q_{\rm prev}$, was 16.47 -- 17.36 -- 18.34~au where the uncertainties of the comet's orbit are taken into account. We use here and in the following text the same three-percentile notation as before.  When including perturbations of all stars, we observe that they lead to a notable larger $q_{\rm prev}$ of the range of 311.7 -- 354.5 -- 436.2~au. Furthermore, they slightly reduced the previous $1/a$ to values between 27.06 -- 27.33 -- 28.03 in units of $10^{-6}$\,au$^{-1}$ which changes the orbital period to almost 7.5\,Myr. 

The case of C/2015\,XY$_1$ represents a group of comets that suffered a strong perturbation about 1.08\,Myr ago from P0506. This has to be taken into account when we follow their dynamics further into the past. But the reader should be warned that the stellar data for the star P0506, while quite precise, might change in the future since the star {\it Gaia} DR3 5571232122386729856 might appear to be a second component of a double system with P0506. Its astrometry has large uncertainties and we do not know its radial velocity or its mass, so we ignore this secondary in our current model and wait for the next {\it Gaia} data release.

\subsubsection{C/2012~LP$_{26}$~(Palomar)} 

In contrary to the previous examples the past dynamic evolution of C/2012~LP$_{26}$ shows that only HD\,7977 plays a significant role as the stellar perturber. The evolution we show here starts from the original GR~orbit  based on 6.4\,yr data arc in a range of heliocentric distances: 10.01 -- 6.536 (perihelion) -- 9.86~au (solution ‘aa’ in the CODE Catalogue). 

During the backward numerical integration, we again notice small perturbations from P0506 and P0417, but generally, without action from HD\,7977 the past dynamical evolution of C/2012~LP$_{26}$ is almost identical to the case when we exclude all stars, see the right plot in Fig.~\ref{fig:2012lp26_past_future_evolution}. In this case, we obtain the previous perihelion spread according to the comet orbit uncertainty as 15.1 -- 15.6 -- 16.3~au. But if we take into account the nominal influence of HD\,7977, we obtain $q_{\rm prev}$ as large as 588 -- 619 -- 647~au. The semimajor axis and the orbital period remain almost unchanged here.

\subsubsection{C/2010~U3 (Boattini)}

This is one of the very long observed Oort spike comets so far, the pre-discovery data going back to November~2005, and currently, the positional data cover about 18\,yr. The comet passed its perihelion at the end of February~2019 at a distance of 8.45\,au from the Sun and is still observable (as of December~2023). Despite the large perihelion distance, some signs of NG~acceleration in the motion of this comet are seen \citep{Kroli-Dones:2023}.
We used here a very precise orbit obtained using preperihelion data in the range of heliocentric distances from 25.7\,au to 11.9\,au (the solution 'sg' in the CODE catalogue).

\begin{figure}
    \centering
    \includegraphics[width=0.36 \textheight]{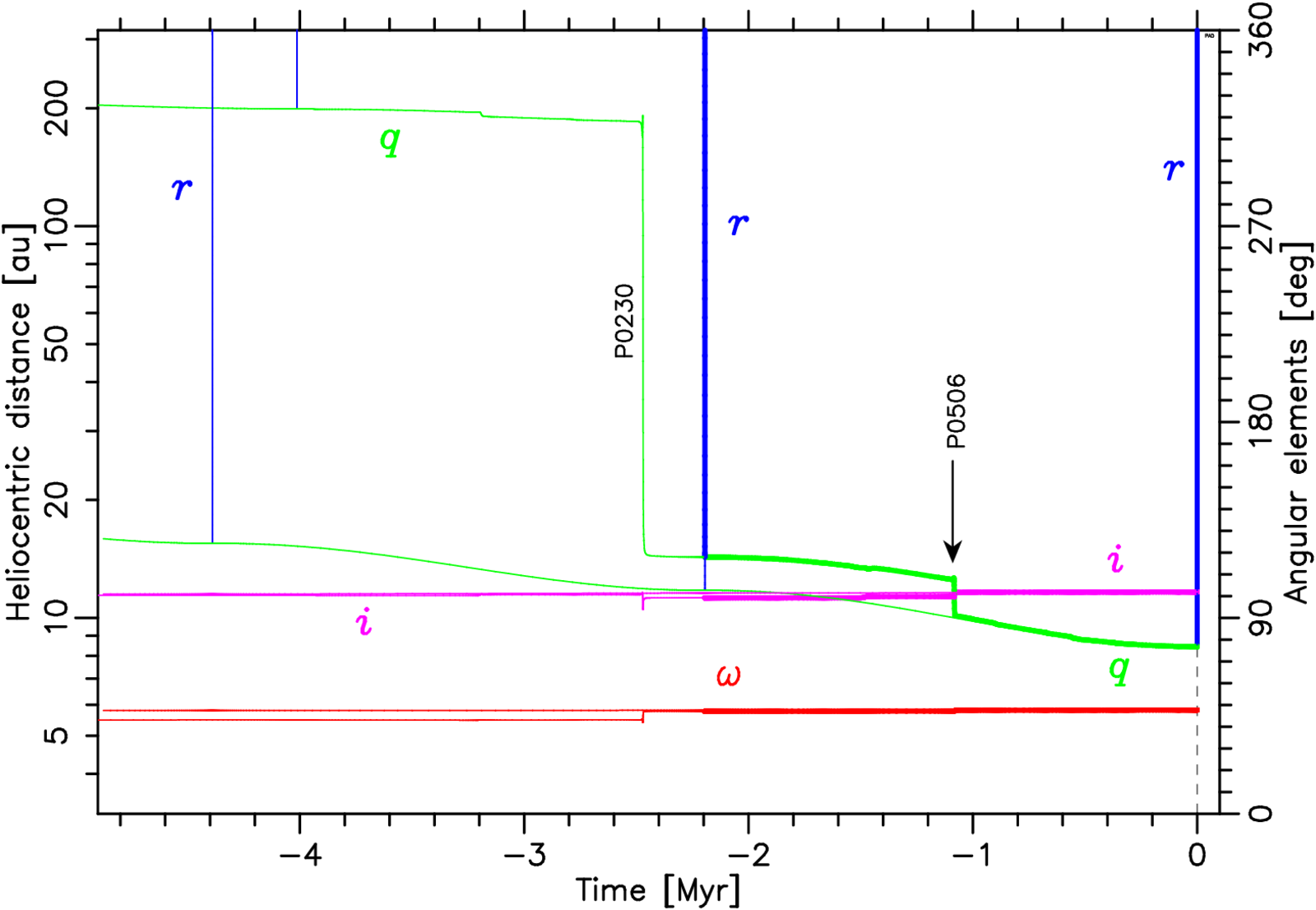}
    \caption{Past dynamical evolution of C/2010~U3 nominal orbit (“sg” solution, see CODE Catalogue). Changes in a perihelion distance (green), an inclination (fuchsia), and an argument of perihelion (red) are shown; angular elements are expressed in a Galactic frame. The thick lines going back in time to the previous perihelion show the result of the full dynamical model while thin lines in this period -- the evolution of elements in the absence of any stellar perturbations, that is, only the galactic perturbations are taken into account. See text for the interpretation of the plot before 2.2\,Myr ago.
}
    \label{fig:2010u3_past_evolution}
\end{figure}

C/2010~U3 represents the case in which the comet was at the previous perihelion distance (2.2\,Myr ago) after HD\,7977's close encounter with the Sun (2.47\,Myr ago), see (Fig.~\ref{fig:2010u3_past_evolution}). Including the comet orbit uncertainties, we obtained $q_{\rm prev}$ equal to 13.92 -- 14.28 -- 14.67~au. According to our calculations, during the last orbital revolution, this comet suffered only a moderate orbit change due to the action of P0506. 

To speculate about the deeper past, we need a fairly fundamental assumption of the neglect of planetary perturbations during this comet's previous perihelion passage. With this assumption, we can see that when C/2010~U3 motion is followed to the one before the last perihelion, it might have been significantly perturbed by HD\,7977 (the almost vertical green line going up the image in the moment of 2.47\,Myr). According to our calculations, the perihelion distance could change from almost 200\,au down to below 15~au as it is shown in Fig.~\ref{fig:2010u3_past_evolution}.

\section{Possible HD\,7977 influence on other Solar System bodies}\label{sec:other-bodies_past_motion}

The HD\,7977 flyby could also have impacted the dynamics of other objects in the outer Solar System than Oort cloud comets. It is worth clarifying here. When investigating the influence of HD\,7977 on comets in Section \ref{sec:LPCs_past_motion}, we followed their motion from the current position backward to calculate the changes in their orbits 2.5\,Myr ago. In this section, to show the level of perturbations caused by a very close stellar passage on other Solar System bodies' motion, we follow them in an opposite direction, starting a calculation before the star approach and from an assumed initial orbital state. 

To investigate the possible gravitational effect on planets and Transneptunian objects (TNOs), we performed a series of simple numerical simulations using the REBOUND \citep{rebound:2012} $N$-body simulator including either the giant planets or the TNOs. The relative uncertainty of the flyby distance is larger than the distance itself, so the star could pass by the system arbitrarily close. Because the effect of the flyby increases when the distance decreases, we decided to focus on the flybys closer than the nominal distance.

\subsection{Action on planets}
The star flyby close to the Sun could have a prominent impact on the orbits of giant planets and, therefore, could affect the stability of the entire Solar System. We used the relative change of the semimajor axis of Neptune to estimate the impact of stellar flyby on planets. It was shown by \cite{Brown-Rein:2022} that the relative changes of the Neptune semimajor axis $\frac{\Delta a}{a}>0.001$ have a significant probability of affecting stability and $\frac{\Delta a}{a}>0.01$ almost always destabilises the Solar System. We can use this criterion to estimate the closest possible distance of the flyby, where the stability of planets is not affected. 

In our simulations, we numerically integrated the orbits of the Sun, HD\,7977, and the four giant planets. We used the nominal trajectory of the passing star but reduced the minimal distance. We used the present-day orbits of planets, but to accommodate for the uncertainty, we repeated the simulation 500~times, shifting the flyby moment by 1/500 of the Neptune orbital period to cover the whole possible range of orbital positions of this planet. Each simulation starts 25,000\,yr before the flyby (the moment when HD\,7977 passes the perihelion of hyperbolic orbit around the Sun) and ends one million years after the flyby. This allows us to estimate not only the direct impact on the dynamics but also the indirect changes of the orbits due to gravitational interactions between planets perturbed by the passing star. Then, at each step of the integration, we compare the orbits of the planets with the orbits resulting from the simulation without the passing. The greatest difference in the semimajor axis during the whole simulation we take as the estimate of flyby impact. 

\begin{figure}
    \centering
    \includegraphics[width=0.4 \textheight]{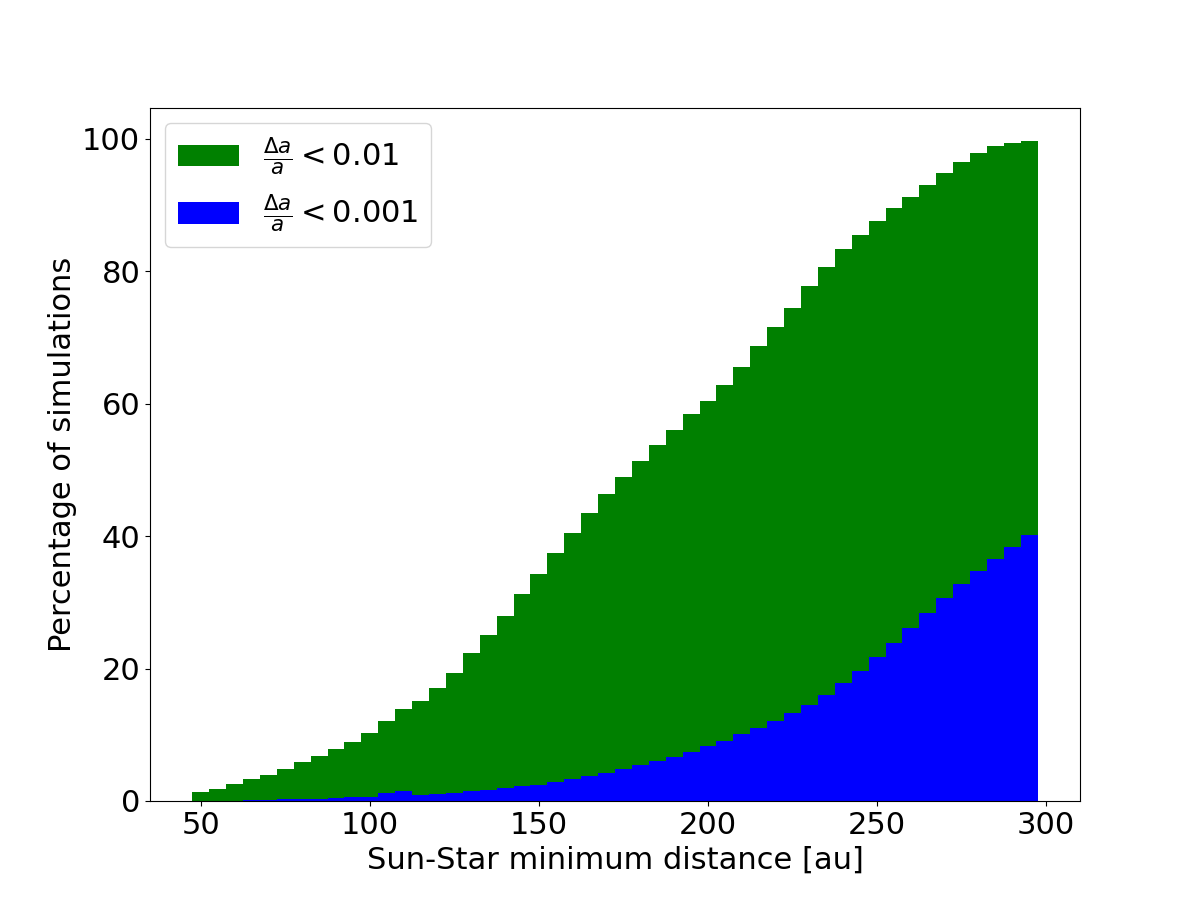}
    \caption{The possible effect of a very close flyby of HD\,7977 on Neptune's semimajor axis. The color bars show the percentage of simulations where the maximum relative difference in a semimajor axis of that planet with and without the flyby is lower than 0.01 or 0.001, respectively.
    }
    \label{fig:neptune_a}
\end{figure}

In Fig.~\ref{fig:neptune_a} we show the percentage of flyby simulations in which the relative change of the Neptune semimajor axis reaches the limits of $\frac{\Delta a}{a}<0.001$ and $\frac{\Delta a}{a}<0.01$ at different distance close flybys. 
The simulation percentage can be used to estimate the probability of orbital elements changing by a certain order of magnitude. For very close flybys, in most cases, the $0.001$ limit is exceeded, but there is still a significant number of simulations in which the change of Neptune's orbit is small enough to maintain the stability of the Solar System. 

\subsection{Action on Transneptunian objects}
The Transneptunian bodies can be divided simply into two subgroups: Kuiper Belt Objects (KBOs) and Scattered Disc Objects (SDOs). The objects in the first group are dynamically cold and have low eccentricities and inclinations. The semimajor axis is bounded by the orbit of Neptune ($\approx 30\text{au}$) and the distance of the mean motion resonance 1:2 with Neptune ($\approx 44\text{au}$). At larger distances, the number of objects with low eccentricity and inclination decreases significantly. The second group (SDOs) is dynamically hot, with higher eccentricities and inclinations. This group is not bounded by 1:2 resonance with Neptune, and some objects have a semimajor axis of hundreds of au. The effect of the flyby is stronger on objects more distant from the Sun, so we focus on the object from the second group.

Instead of simulating the effect of a passing star on the currently observed population of SDOs, we decided to use a more uniform distribution of orbital elements. This allows us to check if a close flyby similar to HD 7977 could possibly alter the distribution of studied objects and reshape this arbitrarily constructed population to be closer to the current one (ex. by increasing the eccentricities in the scattered disc and removing the dynamically cold objects from the outer TNO region).
In our simulations, we randomly generated a population of objects with a semimajor axis greater than 50\,au, an eccentricity less than 0.5, and an inclination smaller than 45~degrees. Then we performed the numerical simulation of the flyby and investigated the orbital elements after the flyby of HD\,7977-like star, i.e. the star with the same mass and velocity as HD\,7977, but we checked different flyby geometries. The change in TNO orbit during flyby comes from two effects. The first is the direct change of the TNO velocity due to the gravitation of the passing star. The second is an indirect effect caused by the change in the velocity of the Sun as a result of the flyby. For the Sun, the total velocity added by the flyby is directed toward the perihelion of the hyperbolic orbit of the passing star and depends on the distance of that perihelion. Similarly, the change in velocity of TNO depends on the minimum distance between the object and the star. Because our simulated population is a disc of objects with low inclination to the ecliptic plane, we can estimate the direct effect of flyby on TNOs by the distance between the Sun and the Star when a passing star crosses the ecliptic plane. That distance is equal to or greater than the perihelion distance and depends on the geometry of the flyby. 

\begin{figure}
    \centering
    \includegraphics[width=0.4 \textheight]{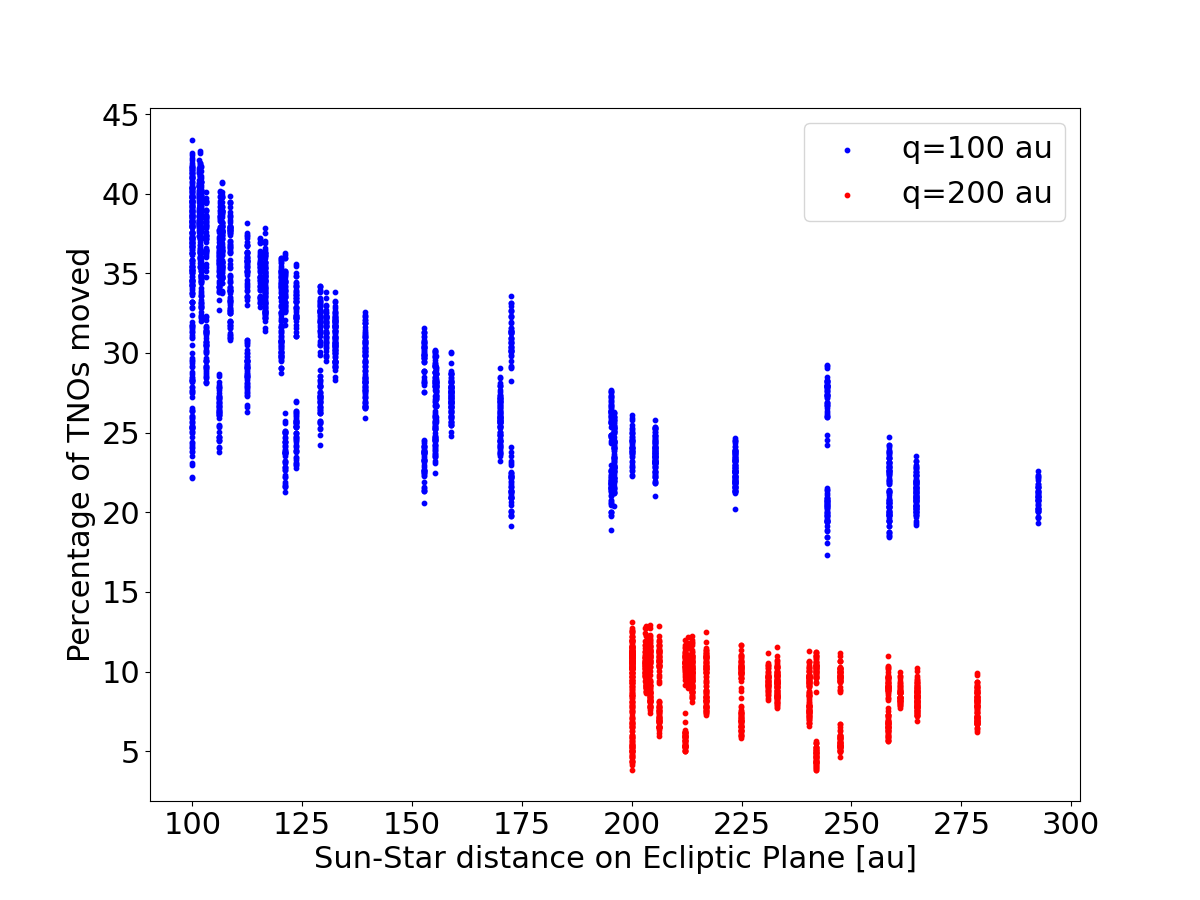}
    \caption{The effect of different flybys on distant TNOs. The plot shows the percentage of simulated objects with an initial semimajor axis between 50 and 100\,au and eccentricity$<0.25$, that have orbital parameters outside this range after a flyby. Each point represents the single simulation of a stellar flyby with a perihelion distance of the Star $q$ equal to 100 or 200 au and different flyby geometry.
    }
    \label{fig:KBO1}
\end{figure}

\begin{figure}
    \centering
    \includegraphics[width=0.4 \textheight]{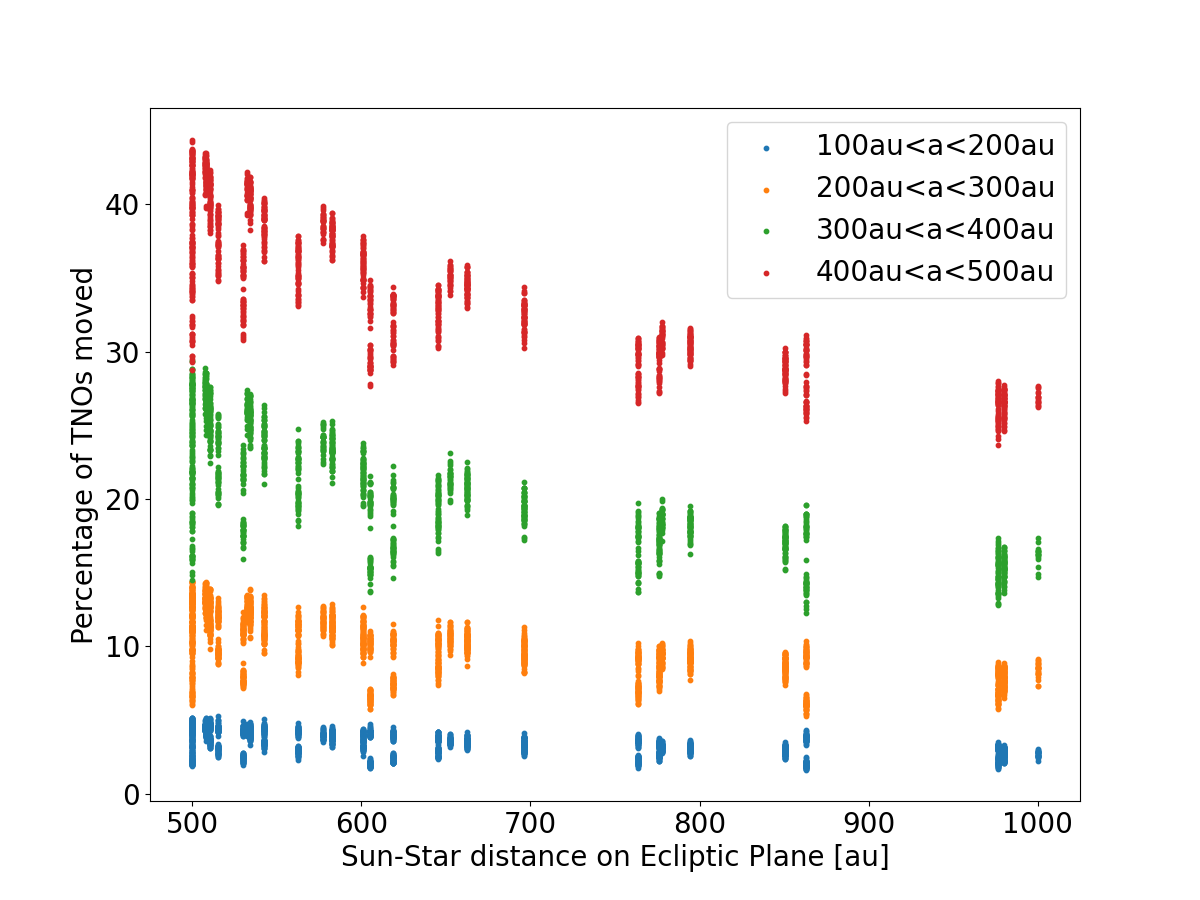}
    \caption{The effect of flybys with perihelion distance 500 au on hypothetical very distant TNOs. The plot shows the percentage of simulated objects with an initial semimajor axis in a given range and eccentricity$<0.25$, that have orbital parameters outside the initial range after a flyby. Each point represents a single simulation and different colour represents a different population of TNOs.
    }
    \label{fig:KBO2}
\end{figure}

In Fig.~\ref{fig:KBO1} we show the effect of a passing star on possible low-to-medium eccentricity objects (initial $e<0.25$) between 50 and 100\,au. The plot shows the percentage of simulated TNOs that were moved by the flyby outside of their initial zone (orbital elements after the flyby are not in the initial range $50<a<100$au and $e<0.25$). 
Each point represents a single simulation, each having a different flyby geometry. Extremely close flybys with perihelion at 100\,au can disrupt 20\% to 45\% objects, and the maximum value of the effect depends on the distance between the Sun and the flying star measured at the moment when the star crosses the ecliptic plane. For the 200\,au flyby, the number of disrupted objects drops to 5-15\%. For flybys farther away with a perihelion distance greater than 300\,au, the percentage of disturbed objects is less than 2\%.  As shown in the previous paragraph, such close encounters (<300\,au) can possibly affect the stability of the planetary system, so it is unlikely that the flyby could explain the low abundance of dynamically cold objects at greater heliocentric distances. If we consider theoretically more distant objects, they can be moved outside their initial zone by a more distant flyby (results for the 500\,au distant flyby shown in Fig.~\ref{fig:KBO2}). 

This simple simulation shows that the overall effect on distant TNOs of an even very close flyby is rather small. Using the fact this effect increases with a semimajor axis, we can extrapolate this result to smaller heliocentric distances and claim that the possible change in orbital elements of closer Kuiper Belt Objects is even smaller. This simple simulation can not however rule out possible indirect orbital effects coming from the change of orbit of major planets due to the flyby, as it is described by \citet{Raymond_et_al:2023}.

Our simulations of a close star passage effect on TNOs shows that the possibility of a very close flyby should not be disregarded using the argument that it must have affected the stability of the Solar System or disrupted the TNOs distribution. However, basing on these simple dynamical tests, we can not set the strict minimum flyby distance for HD 7977 or any other star.

\section{Conclusions and prospects}
\label{sec:conclusions}

The main purpose of this paper is to announce that a very close stellar passage near the Sun cannot be excluded on the basis of current data. From this perspective, the most important star is HD\,7977 but we also indicate for the first time another potential massive perturber, namely HD\,102928. We discuss in detail the quality of all current {\it Gaia} results for HD\,7977 that lead to this finding.  We also tried to check the validity of its current proper motion estimates, the most important parameters from the point of view of the proximity details.

Concluding that these estimates cannot be ruled out, we obtained the closest distance of HD\,7977 $\leq$ 0.012 pc with probability 10\% or $\leq$ 0.060 pc with probability 90\%, which happened 2.47\,Myr ago. We show the influence of such a close passage of one solar mass star on the orbits of long-period comets. We also show with a series of simple simulations what the effect of such a passage on other bodies of the solar system may be. No matter which star made a very close passage near the Sun a few million years ago, this event would constitute some kind of dynamical horizon for all studies of the past dynamics of selected Solar System small bodies, for example, LPCs.

As it concerns LPCs it should be stressed that the long-standing problem of discriminating between dynamically old and new comets now seems to be much more complicated. Past dynamics of almost all LPCs strongly depend on stellar perturbations, but these cannot be currently calculated precisely due to still large stellar data uncertainties and an incomplete list of stars to be taken into account.

The main conclusion regarding the HD\,7977 close passage event is that currently it cannot be ruled out but also cannot be described in detail. We simply have to wait for more precise data from the {\it Gaia} mission, but a dedicated study of the historical positions of this star would also be of great value. Moreover, accurate data for HD\,102928 are very necessary.

Our research also pointed out that there exists another line of study, so far almost absent in the literature, namely the search for resolved binary or multiple systems that have passed or will pass near the Sun. Such events may appear to be more important because of the larger total mass of such a perturber. But all candidates we can find suffer from large uncertainties of their astrometry and/or radial velocity and/or mass estimates. This makes the percentage of multiples in the StePPeD database of potential perturbers much too low. 

\begin{acknowledgements} 
We would like to express our thanks to the anonymous
referee for so many so detailed and so helpful suggestions and comments. These allowed us to greatly improve this paper.
This research has made use of the SIMBAD database, operated at CDS, Strasbourg, France. This work has made use of data from the European Space Agency (ESA) mission {\it Gaia} (\url{https://www.cosmos.esa.int/gaia}), processed by the {\it Gaia}
Data Processing and Analysis Consortium (DPAC,
\url{https://www.cosmos.esa.int/web/gaia/dpac/consortium}). Funding for the DPAC has been provided by national institutions, in particular the institutions
participating in the {\it Gaia} Multilateral Agreement. The calculations that led to this work were partially supported by the project "GAVIP-GC: processing resources for {\it Gaia} data analysis" funded by the European Space Agency (4000120180/17/NL/CBi). P.B. acknowledges funding through the Spanish Government retraining plan 'María Zambrano 2021-2023' at the University of Alicante (ZAMBRANO22-04). E.P-G.: This work has been partially supported
by Horizon 2020 grant no. 871149 ‘EPN-2024-RI. K.L. acknowledges funding by the Italian Space Agency (ASI)
within the Hera agreement (ASI-UNIBO, n. 2022-8-HH.0).

\end{acknowledgements}

\bibliographystyle{aa} 
\bibliography{PAD32} 

\begin{appendix}
\section{Detailed remarks on the HD\,7977 astrometric data in \textit{Gaia}} \label{Appendix-astrometry}

\begin{figure}
    \centering
\includegraphics[width=\columnwidth]{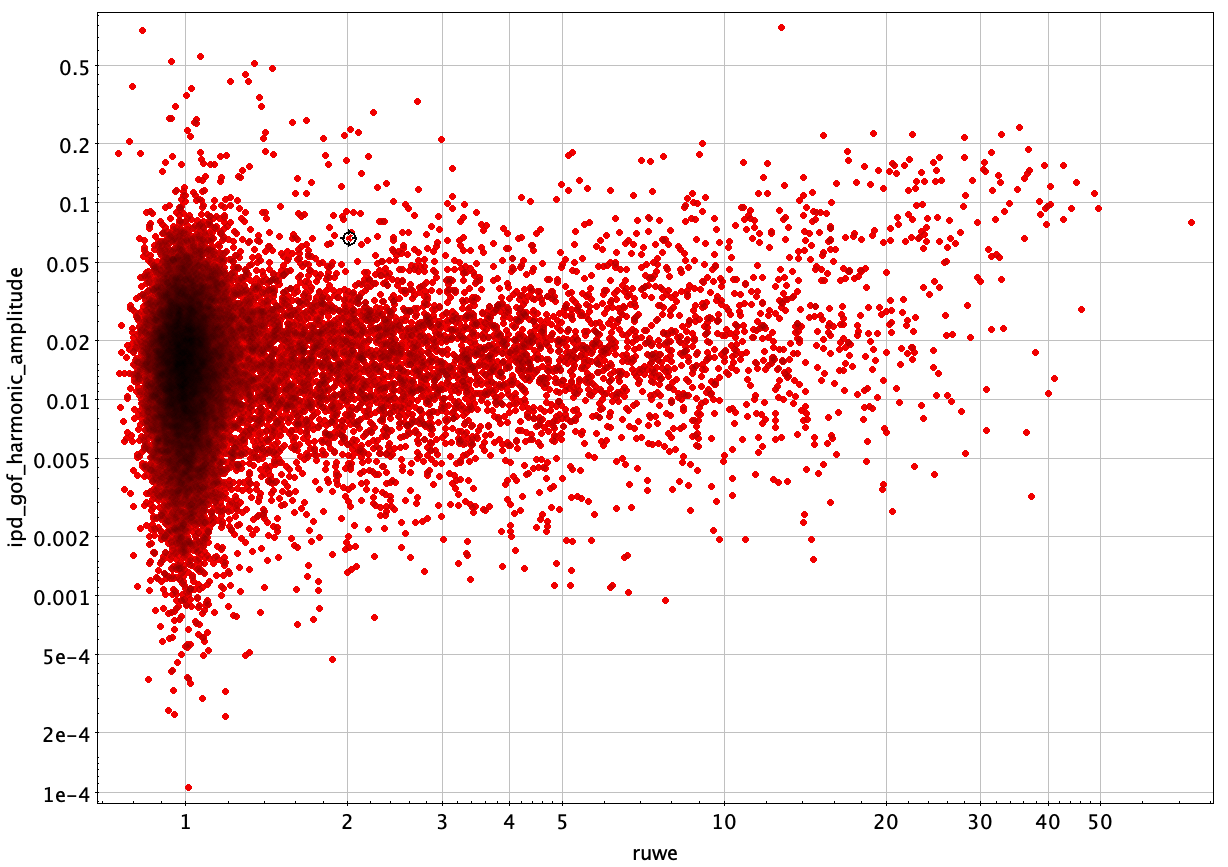}
    \caption{{\it Gaia} DR3 astrometric properties of a comparison sample of 25\,397 stars for HD\,7977: ipd\_gof\_harmonic\_amplitude versus ruwe. The sources have comparable scanning geometry and history (ecliptic latitude within 5~degrees) and similar brightness as HD\,7977 ($G$-band magnitude within 1 magnitude). HD\,7977 is indicated with a cross-hair. It falls clearly in the domain with poor astrometric fits (${\rm ruwe} > 1.25$) and extended source structure (elevated value of ${\rm ipd\_gof\_harmonic\_amplitude}$), compatible with duplicity.}
    \label{fig:RUWE}
\end{figure}

In a comparison sample with 25,397 objects which have a similar scanning geometry and history (ecliptic latitude within 5 degrees) and a similar brightness as HD\,7977 ($G$-band magnitude within 1 magnitude), only 13\% have a larger ruwe value (Fig.~\ref{fig:RUWE}).

An independent suggestion for binarity is given by ${\rm ipd\_gof\_harmonic\_amplitude} = 0.07$ for HD\,7977, which is greater than the value of 97\% of the stars in the comparison sample mentioned above. This harmonic amplitude quantifies the variation of the goodness of the epoch image parameter determination (ipd) fits as a function of the position angle of the scan direction. A nonzero value indicates the presence of a nonisotropic spatial structure of the source, such as caused by a binary that is marginally resolved by {\it Gaia} on selected transits.

Only 1\% of the stars in the comparison sample have a larger ruwe and a larger ipd\_gof\_harmonic\_amplitude value than HD\,7977, providing further evidence that the deviating values are not caused by noise but by duplicity. It should be emphasised that further astrometric data of the object do not indicate anything suspicious that could otherwise explain the high ruwe and/or ipd\_gof\_harmonic\_amplitude values. For example, the 54 epoch observations are well spread over the 34 months of mission data, resulting in 28 visibility periods (“independent observation clusters”) and only 1\% of the 474 detector transits have been excluded from the astrometric fit.

The astrometric excess noise $\epsilon$, in addition to ruwe, provides further quantification of the level of disagreement of the astrometric epoch data with the best-fit single-star motion model. For HD\,7977, $\epsilon = 0.28$~mas. This value has been added in quadrature to the observational epoch uncertainties in order to statistically match the residuals in the astrometric solution. 
In summary, the real error of the astrometry of this particular source exceeds the formal uncertainties reported. Moreover, {\it Gaia} astrometry, in particular of bright stars, is generally known to suffer from underestimated uncertainties. \cite{2021ApJS..254...42B}, for instance, reports the need for inflation of astrometric uncertainty by a factor 1.37 for stars brighter than $\sim$12~mag and other studies have found even larger statistical inflation factors, up to a factor $\sim$2 or more \citep[e.g.,][and references therein]{2021MNRAS.506.2269E,2022A&A...657A.130M}.

Two further remarks can be made on the {\it Gaia} astrometric uncertainties:
\begin{itemize}
\item The five astrometric parameters do not only come with five associated formal errors but with a full, $5 \times 5$ co-variance matrix reflecting the fact that the parameters result from one fit. For example, for the two proper motion components, the reported correlation coefficient $\rho = -0.36$ such that a positive random error in the right ascension proper motion component would be accompanied by a negative random error in the declination proper motion component.
\item \cite{2021A&A...649A.124C} provide convincing evidence that the {\it Gaia} DR3 reference frame suffers from a magnitude-dependent spin. These authors also provide a global fit, as a function of magnitude, and an associated correction formula. For HD\,7977 (${\rm ra} = 20.131652206655247$ deg, ${\rm dec} = 61.882521645503466$ deg, $G = 8.891319$ mag), this correction suggests that the proper motions of {\it Gaia} in this direction in the sky should be corrected by adding $0.031$ and $0.025$~mas~yr$^{-1}$ in right ascension and declination, respectively (despite acknowledging that colour-dependent systematics of up to $\sim$$0.010$~mas~yr$^{-1}$ can remain present). These corrections are of the same order of magnitude as the formal proper motion uncertainties (0.024 and 0.034~mas~yr$^{-1}$).
\end{itemize}
...
\section{Other HD\,7977 distance estimations found in the literature.} \label{Appendix-distances}

Several other distance estimates are available. However, since the formal parallax error is so small, these are highly correlated with the inverse-parallax estimate reported above:
\begin{itemize}
\item \cite{2021AJ....161..147B} present Bayesian distance estimates for stars in {\it Gaia}~DR3. The first, dubbed geometric distance, uses the measured parallax with a direction-dependent prior to the distance and results in $75.546 \pm 0.195$\,pc. The second, dubbed photo-geometric distance, uses the colour and apparent magnitude in addition and results in $75.560 \pm 0.163$~pc. In both cases, the distance estimate is dominated by the high signal-to-noise measured parallax and only marginally influenced by the prior.
\item {\it Gaia} DR3 itself contains a Bayesian distance estimate, dubbed distance\_gspphot, returned by the General Stellar Parameterizer from Photometry
\citep[GSP-Phot;][]{Andrae_et_al:2023}, which is part of the astrophysical parameter inference system
\citep[Apsis;][]{2023A&A...674A..26C}. GSP-Phot simultaneously fits the low-resolution BP/RP spectrum, the measured parallax, and the apparent $G$ magnitude and returns a distance estimate of $75.791-0.401+2.846$~pc and an absolute magnitude $M_G = {\rm mg\_gspphot} = 4.47$~mag (with an uncertainty interval [4.35, 4.49]~mag) using ${\rm ag\_gspphot} = 0.005$~mag. Again, most of the weight in this distance estimate comes from the high-precision parallax measurement.
\item \cite{2022A&A...658A..91A} also present Bayesian distance estimates for stars in {\it Gaia} DR3 after adding photometric constraints from the Pan-STARRS1, SkyMapper, 2MASS, and AllWISE catalogues. The associated StarHorse code returns a distance estimate of $75.437\pm0.402$~pc and an absolute magnitude $M_G = 4.31 \pm 0.10$~mag using $A_G = 0.20 \pm 0.10$~mag. Again, the prior in this distance estimate has negligible influence because of the small formal uncertainty on the measured parallax. The addition of the (infrared) photometry has shifted $\sim$0.2~mag between absolute magnitude and $G$-band extinction between the StarHorse and the GSP-Phot estimates, indicative of some ill-defined properties of the source possibly linked to unrecognised duplicity.
\end{itemize}
It is important to recall that all above distance estimates implicitly assume that the source is a single one.

\section{Detailed remarks on the HD\,7977 \textit{Gaia} spectral data.}
\label{Appendix-spectra}
\begin{figure}
    \centering
\includegraphics[width=\columnwidth]{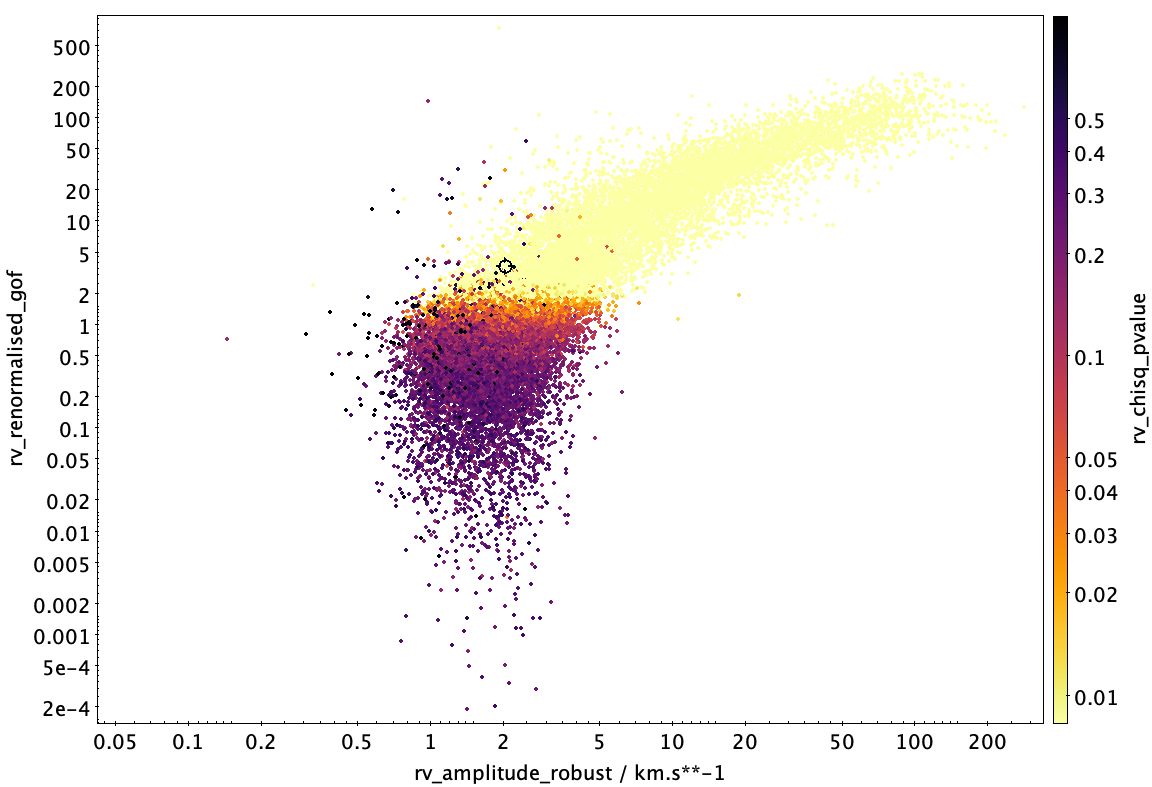}
    \caption{{\it Gaia} DR3 RVS properties of a comparison sample of 35,629 stars for HD\,7977: rv\_renormalised\_gof versus rv\_amplitude\_robust, colour-coded with rv\_chisq\_pvalue. The sources have comparable spectral types (dwarfs with a template effective temperature within 500\,K) and similar brightness as HD\,7977 ($G_{\rm RVS}$-band magnitude within 1~magnitude). HD\,7977 is indicated with a cross-hair. It falls clearly in the domain where the radial velocity time series is not constant (${\rm rv\_chisq\_pvalue} < 1$E-2) and also the normalised goodness of fit is excessively large (compared to unity).}
    \label{fig:RVS}
\end{figure}

{\it Gaia} DR3 provides a measurement of the peak-to-peak amplitude of the time series of radial velocity after filtering out “probable outliers” \citep[see Section 3.7 in][]{2023A&A...674A...5K}. This scatter equals ${\rm rv\_amplitude\_robust} = 2.04$~km~s$^{-1}$ with an associated ${\rm rv\_time\_duration} = 911$~days (3.4\,years). The long-term scatter being an order of magnitude larger than the formal uncertainty on the mean radial velocity is a suggestive indication of radial-velocity variability. To quantify this further, the renormalised goodness of fit, rv\_renormalised\_gof, is a quantity computed using the source under inspection and all other sources having similar rv\_template\_teff and grvs\_mag such that the scatter of the epoch radial velocities of the source can be meaningfully compared to the typical epoch uncertainty for the appropriate grvs\_mag and rv\_template\_teff ranges. For HD\,7977, we find ${\rm rv\_renormalised\_gof} = 3.59$, which is clearly anomalous (for constant stars, rv\_renormalised\_gof follows a normal distribution with a standard deviation equal to one). In a comparison sample with 35,629 objects which have comparable spectral types (dwarfs with a template effective temperature within 500~K) and similar brightness as HD\,7977 ($G_{\rm RVS}$-band magnitude within 1 magnitude), only 20\% have a larger rv\_renornalised\_gof value (Fig.~\ref{fig:RVS}).

To find (potential) radial-velocity variable stars, the following selection is suggested in Section~11 in \cite{2023A&A...674A...5K}: ${\rm rv\_chisq\_pvalue} < 1$E-2  and ${\rm rv\_renormalised\_gof} > 4$  and ${\rm rv\_nb\_transits} \ge 10$. For HD\,7977, two of these three conditions are met with margin (${\rm rv\_chisq\_pvalue} = 2$E-6 $<$ 1E-2 and ${\rm rv\_nb\_transits} = 18 \ge 10$) with only the renormalised goodness of fit ${\rm rv\_renormalised\_gof} = 3.59$ staying slightly below the formal limit of $4$.

In conclusion, the (unpublished) radial velocity time series contains hints for the presence of variability, most likely caused by duplicity, but further quantification and definitive conclusions require the publication of {\it Gaia}~DR4. 

The {\it Gaia} low-resolution blue and red photometer spectra (BP and RP) and medium-resolution radial velocity spectrometer (RVS) spectra provide significant additional information on the astrophysical characteristics of the sources. There are several (inconsistent) effective temperature estimates for HD\,7977:
\begin{itemize}
\item ${\rm teff\_gspphot} = 5816$~K with an uncertainty interval $[5807, 5894]$~K based on the BP/RP spectra, compatible with a G1V/G2V spectral type (compared to a G1V spectral type inferred from the GSP-Phot absolute magnitude);
\item ${\rm teff\_gspspec} = 5956$~K with an uncertainty interval $[5924, 6048]$~K based on the RVS spectrum, compatible with a G0V spectral type;
\item $6149$~K with an uncertainty interval $[5938, 6280]$~K derived by \cite{2022A&A...658A..91A} with the StarHorse code that uses {\it Gaia} DR3 data and additional photometric constraints, compatible with an F8V spectral type (compared to a G0V spectral type inferred from the StarHorse absolute magnitude).
\end{itemize}
For reference, the nominal spectral type in {\it SIMBAD} is G3 (without luminosity class).

For HD\,7977, the BP/RP provides hints on possible unresolved binarity through the discrete source classification \citep[DSC,][]{Delchambre_et_al:2023} and the multiple star classification \citep[MSC,][]{2023A&A...674A..26C}.
The DSC probability of being a binary star, derived from an ExtraTrees classifier that uses the BP/RP spectrum \citep[][Section~11.3.2]{2022gdr3.reptE..11U}, equals ${\rm classprob\_dsc\_specmod\_binarystar} = 0.19$, with the complementary probability ${\rm classprob\_dsc\_specmod\_star} = 0.81$. Obviously, the binary star probability is below 50\%, and, in addition, the DSC results for binaries should be used with caution \citep[][Section~11.4.4]{2022gdr3.reptE..11U}. Nonetheless, the finite value of classprob\_dsc\_specmod\_binarystar does hint at a spectral peculiarity consistent with unresolved duplicity.

The MSC assumes that the BP/RP spectrum is a composite spectrum of an unresolved (coeval) binary with shared metallicity, extinction, and distance \citep[][Section~11.3.5]{2022gdr3.reptE..11U}. It returns these common parameters as well as the effective temperature and surface gravity for both components: $T_{\rm eff,1} = 6099$~K with an uncertainty interval $[6084, 6112]$~K and $T_{\rm eff,2} = 5198$~K with an uncertainty interval $[5134, 5279]$~K, with an estimated extinction of the $G$ band of $0.17 \pm 0.04$~ mag and an estimated distance of $97 \pm 1$~pc. 
This distance estimate deviates significantly from the (single-star) distance estimates of $\sim$75\,pc (see above) and also the extinction differs significantly from the GSP-Phot value, ${\rm ag\_gspphot} = 0.005$~ mag (uncertainty interval $[0.0005, 0.0472]$~ mag), but agrees with the StarHorse estimate of $A_G = 0.20 \pm 0.10$~mag.

All in all, the {\it Gaia} spectra do not provide unique constraints on the astrophysical nature of the source and leave the possibility of an unresolved binary open.

\end{appendix}

\end{document}